\documentclass[journal=jpcafh,manuscript=article]{achemso}
\usepackage{longtable}
\usepackage{multirow}
\usepackage{amsmath}
\usepackage{subfigure}
\usepackage{cleveref}
\usepackage[usenames,dvipsnames]{color}
\usepackage{soul}
\usepackage{bm}
\usepackage{rotating}
\usepackage{amssymb}
\usepackage{titlecaps}
\usepackage{xcolor}
\usepackage{soul}
\newcommand{\norm}[1]{\left\lVert#1\right\rVert}
\Addlcwords{a an the of on at with for and in into from by without}
\usepackage{units}
\usepackage{tikz}
\usepackage{import}
\usepackage[normalem]{ulem}
\usepackage[sort&compress,numbers,super]{natbib}
\mciteErrorOnUnknownfalse
\usepackage{xr}
\usepackage{multicol} \usepackage{wrapfig} \usepackage{enumitem}
\usepackage{bm} \usepackage{outlines} 

\SectionNumbersOn

\author{Luis Itza Vazquez-Salazar} \affiliation[University of
  Basel]{Department of Chemistry, University of Basel,
  Klingelbergstrasse 80 , CH-4056 Basel,
  Switzerland.}\altaffiliation{These authors contributed equally}
\email{luisitza.vazquezsalazar@unibas.ch}

\author{Silvan K\"aser} \affiliation[University of Basel]{Department
  of Chemistry, University of Basel, Klingelbergstrasse 80 , CH-4056
  Basel, Switzerland.}\altaffiliation{These authors contributed
  equally}\email{silvan.kaeser@unibas.ch}

\author{Markus Meuwly} \affiliation[University of Basel]{Department of
  Chemistry, University of Basel, Klingelbergstrasse 80 , CH-4056
  Basel, Switzerland.}  \email{m.meuwly@unibas.ch}

\title{Outlier-Detection for Reactive Machine Learned Potential
  Energy Surfaces}

\begin{document}
\date{\today}

\begin{abstract}
Uncertainty quantification (UQ) to detect samples with large expected
errors (outliers) is applied to reactive molecular potential energy
surfaces (PESs). Three methods - Ensembles, Deep Evidential
Regression (DER), and Gaussian Mixture Models (GMM) - were applied to the
H-transfer reaction between {\it syn-}Criegee and vinyl
hydroxyperoxide. The results indicate that ensemble models provide the
best results for detecting outliers, followed by GMM. For example,
from a pool of 1000 structures with the largest uncertainty, the detection
quality for outliers is $\sim 90$ \% and $\sim 50$ \%, respectively,
if 25 or 1000 structures with large errors are sought. On the contrary, the limitations of the statistical assumptions of DER greatly impacted its prediction capabilities. Finally, a
structure-based indicator was found to be correlated with large
average error, which may help to rapidly classify new structures into
those that provide an advantage for refining the neural network. 
\end{abstract}

\section{Introduction}
Detecting infrequent and/or out-of-distribution events is
central to data-driven research. Fields in which such phenomena are relevant range from finance\cite{vilalta:2002} to medicine,\cite{lazarevic:2004} 
climate,\cite{frei:2001} weather and the natural sciences.\cite{voter:1997} While ``expected''
outcomes can be typically sampled from a known, computable and controllable
distribution, infrequent (or "rare") events can not always be easily associated with a
predetermined distribution. In most cases it is, however,
the rare events that profoundly affect the development of a system,
such as a crash in stock markets, a tornado in weather, or a bond
breaking/forming process in chemistry. A typical chemical bond with a
stabilization energy of $\sim 20$ kcal/mol (equivalent to a lifetime of 1
s$^{-1}$) and a vibrational frequency of 20 fs$^{-1}$ vibrates $\sim
10^{13}$ times before breaking which makes chemical reactions a ``rare
event''. As the energy in the system increases for
bond breaking (and bond formation) to occur, the available phase space increases
in concert and sampling all necessary regions becomes a daunting
task.\\

Computer simulations are an indispensable part of today's research and
have become increasingly important in chemistry, physics, biology and
materials science. One particularly fruitful approach for the chemical
and biological sciences are molecular dynamics (MD)
simulations\cite{van1990computer,ferrario2006computer,cui2016perspective,raabe1998computational}
that involve the numerical integration of Newton's equations of
motion. This requires the knowledge of the underlying intermolecular
interactions (the ``potential energy surface'' (PES)) and forces
derived from them for a given atomic configuration ${\bf
  x}$.\cite{unke2021machine,behler2021four} Ideally, those properties
would be determined at the highest level of accuracy by solving the
time-independent Schr\"odinger equation (SE). Unfortunately, this is only
possible for small systems on a short time scale because the methods
to solve the SE scale poorly with the system size and the method's
accuracy. This limitation can be circumvented by using atomistic
potentials that directly describe the relation between the atomic
positions of a molecule and its potential energy through the mapping,
$f:\{Z_{i},{\bf x}_{i}\}^{N}_{i=1}\rightarrow E({\bf x})$, of the
atomic charges ($Z_{i}$) and the atomic positions (${\bf x}_{i}$) to
the potential energy $E({\bf x})$ from which the forces can be
determined from the potential energy as its negative gradient ($F_{i}
= -\nabla E({\bf x})$). \\

\noindent
Over the last decade, machine learning (ML) techniques such as neural
networks (NNs) and kernel methods have been used to represent
PESs.\cite{chmiela:2017,unke2021machine,manzhos:2020,MM.rkhs:2017,jiang:2016}
This originates from the methods' ability to \textit{learn}
relationships from data.\cite{MM.rev:2023} Therefore, it is possible
to parametrize/learn the described mapping from a pool of reference
\textit{ab initio} calculations and eventually use it for following
the dynamics of a system of interest. Particularly, ML has been
extensively used to represent PESs based on large, diverse, and
high-quality electronic structure
data.\cite{wang2008full,tew2016ab,kidwell2016unimolecular,li2020many,mm.fad:2022,MM.criegee:2023,MM.morphing:2024}
While Machined Learned Potential Energy Surfaces (ML-PESs), sometimes
also called ML potentials\footnote{Although in the literature it is
  common to find both names, the present work uses ML-PES to avoid
  confusion with multilayer perceptron also known as MLP.} (MLP),
reach remarkable accuracies (orders of magnitude better than
``chemical accuracy'', i.e. 1 kcal/mol) in the interpolation regime of
the data set they are known to extrapolate poorly on unseen data due
to their purely mathematical nature lacking any underlying functional
form.\cite{haley1992extrapolation,behler2011neural} Thus, ML-PESs
crucially depend on the \textit{globality} of the training data, which
usually requires an iterative collection/extension of a data
set.\cite{unke2021machine,MM.rev:2023,tokita2023train}\\

\noindent
On the other hand, constructing globally valid ML-PESs in particular
for chemical reactions is still a challenging task because the phase
space that needs to be covered increases exponentially with the energy
that is required to drive a conventional chemical reaction. This is
directly related to the quality, completeness and coverage of the data
set used to train the ML algorithm, in particular for NN-based
representations. One way to tackle these critical aspects is through
the use of uncertainty quantification (UQ) with the primary goal of
detecting uncovered regions. Those regions are characterized by the presence of outliers (i.e. samples with largely different behaviour than the other members of the dataset\cite{grubbs1969procedures}) which usually have large errors.  Finding such outliers or outlier regions
helps to increase the model's robustness and further improves its
accuracy and reliability. Particularly for reactive PESs - one of the
hallmark applications of ML-based PESs - quantitatively characterizing
the confidence in predicted energies and/or forces for chemically
interesting regions around the transition state(s) (TS) is very
valuable. Such information can be used to distinguish well-covered
regions from those that require additional training data.\\

\noindent
For chemical applications, different UQ techniques have been used. Common are ensemble methods for which multiple independently trained statistical models are used to obtain the average and variance of an observation\cite{gastegger2020molecular}. Depending on the number of ensemble members, their disadvantage lies in the high computational cost they incur. Alternatively, methods based on Gaussian process regression\cite{deringer2021gaussian} were employed, which, however, are limited by the database size for which they can be used.  Alternatives based on single-network methods with the possibility to predict the variance have been proposed, including regression prior networks\cite{malinin2020regression}, mean variance estimation, or Deep Evidential Regression (DER)\cite{amini2020,vazquez2022uncertainty}. The use of some of those methods has been recently benchmarked for non-reactive PESs.\cite{tan2023single} \\

Here the goal is to quantify uncertainty for a reactive system for
which one of the Criegee Intermediates (CIs), {\it syn-}Criegee
(CH$_3$CHOO), was used. The manuscript is structured as
follows. First, the methods, including data set generation,
uncertainty quantification and analysis techniques, are
described. Next, the performance of the PESs for computing geometrical
and energetic properties is assessed. This is followed by the results
on uncertainty quantification, outlier detection and an analysis of
the relationship between molecular structure and
errors/uncertainties. Finally, the findings are discussed in a broader
context and conclusions are drawn.\\

\section{Methods}
This section describes the \textit{ab initio} reference data, the
approaches to quantify uncertainty and further analyses. For the
ensemble and deep evidential regression models, the variance is used
for UQ, whereas 
the negative log-likelihood (NLL) is used for the Gaussian mixture model (GMM).
The ``error'' is the
difference between the reference value of a property and the predicted
value of that property with a given model whereas the ``variance''
defined as the expected value for the squared difference between the
predicted value and the mean value of the model. Finally, uncertainty
is considered as the degree of confidence in the prediction made by a
given model. Uncertainty is related to the lack of knowledge or the
model's limitations to describe a
system.\cite{hullermeier2021aleatoric} In the text, "uncertainty" and
"variance" are used synonymously, whereby a small variance value
corresponds to a smaller uncertainty and a higher confidence in the
prediction and {\it vice versa}. The models are characterized in terms
of the Mean Squared Error (MSE), the Mean Absolute Error (MAE) and the
Mean Variance (MV).\\

\subsection{Data sets}
\begin{figure}
    \centering
    \includegraphics[scale=1.1]{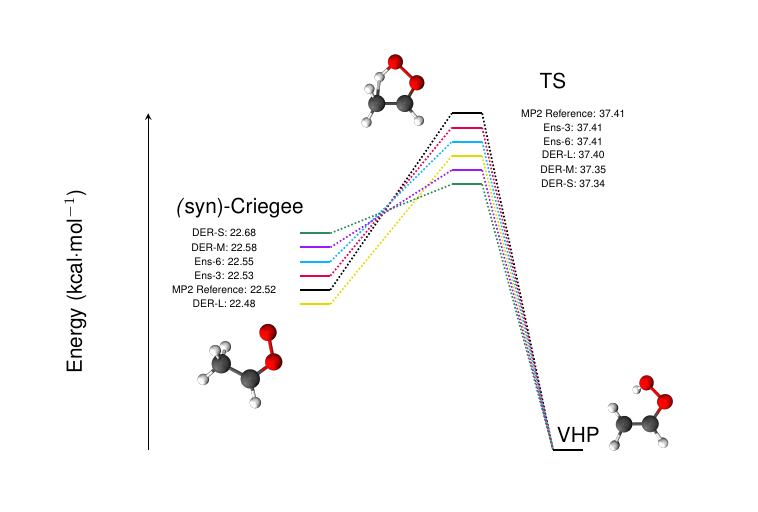}
    \caption{Characteristics of the stationary points of the PESs. The
      energy of the VHP minimum serves as a reference. The energy
      scale is exaggerated to better represent the differences between
      the methods.}
    \label{fig:stat_points}
\end{figure}

The main ingredient for generating ML-PESs is reference electronic
structure data to train the models on. Here, the H-transfer reaction
from (\textit{syn})-Criegee to vinyl hydroxyperoxide (VHP) serves as a
benchmark system (see Figure~\ref{fig:stat_points}) and reference data
at the MP2/aug-cc-pVTZ level of theory is available from previous
work.\cite{upadhyay2021thermal} From a total of 37399 structures
covering the H-transfer reaction for the {\it syn-}Criegee
$\leftrightarrow$ TS $\leftrightarrow$ VHP reaction, $\sim 10$~\% were
extracted semi-randomly (every 10th) and structures with very large
energies ($>400$~kcal/mol above the minimum) are excluded. A total of
3706 data points were used for obtaining a first-generation ML-PES
(see the energy distribution in
Figure~\ref{sifig:gen1_data_energydistr}). Multiple rounds of diffusion
Monte Carlo (DMC) simulations\cite{kosztin1996introduction} and
adaptive sampling\cite{csanyi2004learn} were run to detect
\textit{holes} and under-sampled regions. The resulting final data set
contains a total of 4305 structures (see the energy distribution in
Figure~\ref{sifig:gen3_data_energydistr}) and is used to train new
ML-PESs that are finally used for uncertainty prediction. It is
important to note that the training data set is not considered to be
comprehensive. If, e.g., a global PES for dissociation dynamics
(i.e. formation of vinoxy radical, etc) is sought after, additional
sampling would be required. Nevertheless, the small data set can be
used to obtain different ML-based models and covers the relevant part
of the configurational space of the reactive process of interest
(H-transfer), and their ability to quantify uncertainty can be tested
on an extensive test set. The (unseen) test set contains a total of
33402 structures covering the (\textit{syn})-Criegee $\leftrightarrow$ VHP
reaction and the distribution of energies is shown in
Figure~\ref{sifig:remaining_data_energydistr}.\\

\subsection{Uncertainty Quantification}

\paragraph{Ensembles}
The ensemble method based on the
Query-by-committee\cite{seung1992query} strategy is a frequently used
and practical approach to uncertainty estimation. For this strategy, a
"committee" of models is trained on the same data set. The uncertainty
measure is obtained as the disagreement between the models (or within
the committee/ensemble). If the predictions of the ensemble members
agree closely, it can be assumed that the region on the PES is well
described. For under-sampled regions, however, the predictions will
diverge.\cite{gastegger2020molecular} A commonly used uncertainty
measure for the ensemble is the standard deviation given
by\cite{gastegger2020molecular}
\begin{align}
    \sigma_E = \sqrt{\frac{1}{\mathcal{N}} \sum^\mathcal{N}_n
      \left(\Tilde{E}_n - \Bar{E}\right)^2}.
\end{align}
Here, $\mathcal{N}$ corresponds to the number of committee models,
$\Tilde{E}_n$ is the energy predicted by committee model $n$ and
$\Bar{E}$ is the ensemble average.\\

\noindent
PhysNet\cite{MM.physnet:2019} was chosen to learn a representation of
the PES. A total of 6 models were trained to generate an ensemble. All
models share the same architecture and hyperparameters. However, the
random initialization prior to training and the splits of the
training/validation data were altered (models 1/2, 3/4 and 5/6 were
trained on exactly the same data). The 4305 data points were split
into training/validation sets according to 80/20~\%. The PhysNet
models were trained on energies, forces and dipole moments; see
supporting information. Query-by-committee was performed with an
ensemble of 6 models (Ens-6) and 3 models (Ens-3, models 1, 3, 5).\\

\paragraph{Deep Evidential Regression}
The present work employs a modified
architecture\cite{vazquez2022uncertainty} of PhysNet to predict
energies and uncertainties based on Deep Evidential Regression
(DER). DER assumes that the energies are Gaussian-distributed
$P(E)=\mathcal{N}(\mu,\sigma^{2})$. The prior distribution is a
Normal-Inverse Gamma (NIG), described by four values ($\gamma$, $\nu$,
$\alpha$, $\beta$).\cite{amini2020} The total loss function
$\mathcal{L}$ includes the NLL,
$\mathcal{L}^{NLL}(x)$, which is regularized by the $\lambda-$scaled MSE, $\mathcal{L}^{R}(x)$, that minimizes the
evidence of incorrect predictions together with energies, forces,
charges and dipole moments for all structures in the training set
\begin{equation}
\begin{aligned}
    \mathcal{L} = \mathcal{L}^{NLL}(E_{\rm ref},E_{\rm pred}) +\lambda(\mathcal{L}^{R}(E_{\rm ref},E_{\rm pred})
    -\varepsilon)  + W_{F}|F_{{\rm
        ref}}-F_{\rm pred}| \\ + W_{Q} \left| Q_{\rm ref} - Q_{\rm pred} \right| + W_{D} \left| D_{\rm ref}-D_{\rm pred} \right|.
\end{aligned}
    \label{eq:forces_der}
\end{equation}
The NN is trained to minimize the difference between the NIG
distribution and $p(E)$. The values of the hyperparameters were $W_{F}
= 52.9177$ \AA\//eV, $W_{Q} = 14.3996$ $e^{-1}$, and $W_{D} = 27.2113$
$D^{-1}$, respectively,\cite{MM.physnet:2019} and $\lambda = 0.15$ and
$\varepsilon = 10^{-4}$ throughout. Note that the forces and dipole
moments were calculated as in the original version of PhysNet. As a
consequence, the variance of the forces can not be obtained because
the derivative of the variance is the covariance matrix between energy
and forces.\cite{klicpera2020fast} This model is referred to as DER-Simple (DER-S).\\

\noindent
\paragraph{Modified Deep Evidential Regression}  The
  effectiveness in predicting uncertainties by DER-S has been recently
  questioned\cite{meinert2023unreasonable,oh2022improving}: Firstly,
  minimizing a loss function similar to Equation \ref{eq:forces_der}
  is insufficient to uniquely determine the parameters of the NIG
  distribution because $\mathcal{L}^{NLL}(E_{\rm ref},E_{\rm pred})$
  is optimized independently of the
  data.\cite{meinert2023unreasonable} This leads to large uncertainty
  in poorly sampled regions.  Secondly, it was shown that optimizing
  $\mathcal{L}^{NLL}(E_{\rm ref},E_{\rm pred})$ is insufficient to
  obtain faithful predictions. Adding the term
  $\lambda(\mathcal{L}^{R}(E_{\rm ref}, E_{\rm pred}) -\varepsilon)$
  as a regularizer addresses this problem but can lead to a gradient
  conflict between the two terms\cite{oh2022improving}.\\

\noindent
Two modifications to DER-S were considered. First, the multivariate
generalization, DER-M, following the work of Meinert and
Lavin\cite{meinert2021multivariate} was implemented. In DER-M, the NIG
is replaced by a Normal Inverse Wishart (NIW) distribution, which is
the multidimensional generalization of the NIG distribution to predict
a multidimensional distribution of energies ($E$) and charges
($Q$). The loss function for DER-M is
\begin{align}
\mathcal{L}  = & \log\left(\dfrac{\nu+1}{\nu-1}\right) - \nu \sum_{j} \ell_{j} +  \frac{\nu+1}{2} \log\left(\det\left(\mathbf{L}\mathbf{L}^{\top} +
    \frac{1}{1+\nu}\mathbf{Y} \cdot
    \mathbf{Y}^{\top}\right)\right)  +\\
&W_{F}\left|F_{\rm pred}-F_{\rm ref} \right| + W_{D} \left| D_{\rm pred}-D_{\rm ref}
    \right| \nonumber
\label{eq:md_der}
\end{align}
where $\mathbf{Y} = [E_{\rm ref},Q_{\rm
    ref}]^{\top}-[\mu_{0},\mu_{1}]^{\top}$.  $\mu_{0}$ is the
predicted energy ($E_{\rm pred}$) and $\mu_{1}$ the respective
predicted total charge ($Q_{\rm pred}$).  Then, the model output
contains six values: the objective values ($E_{\rm pred},Q_{\rm
  pred}$), the corresponding parameters of the covariance matrix
$\mathbf{L}$, $\vec{l} = diag(\mathbf{L})$, and a parameter $\nu$. The
outputs of the model were transformed to become the parameters of the
multidimensional evidential distribution. Details on the construction
of the $\mathbf{L}$ matrix, boundaries of $\nu$ and the uncertainty
are given in the SI.\\

\noindent
For the second modified architecture, a Lipschitz-modified loss
function $\mathcal{L}^{Lips}$ was used\cite{oh2022improving} as a
complementary regularization to the NLL loss
\begin{equation}
    \begin{aligned}
          \mathcal{L}= \mathcal{L}^{NLL}(E_{\rm ref},E_{\rm pred})
          +\lambda(\mathcal{L}^{R}(E_{\rm ref},E_{\rm
            pred})-\varepsilon) + \mathcal{L}^{Lips.}(E_{\rm
            ref},E_{\rm pred}) \\ + W_{F}\left|F_{\rm ref}-F_{\rm
            pred} \right| + W_{Q} \left| Q_{\rm ref}- Q_{\rm pred}
          \right| + W_{D} \left|D_{ \rm ref}-D_{\rm pred} \right|
    \end{aligned}
    \label{eq:lf_lipz}
\end{equation}
Here, $\mathcal{L}^{Lips.}(E_{\rm ref},E_{\rm pred})$ is defined as
\begin{equation}
    \mathcal{L}^{Lips.}(E_{\rm ref},E_{\rm pred}) = \left\{
    \begin{array}{ll}
         (E_{\rm ref}-E_{\rm pred})^{2}& {\rm If} \ \lambda^{2} <
      U_{\nu,\alpha} \\ 2\sqrt{U_{\nu,\alpha}}|E_{\rm ref}-E_{\rm
        pred}|-U_{\nu,\alpha}& {\rm If} \ \lambda^{2} \geq
      U_{\nu,\alpha}
    \end{array}
    \right.
\end{equation}
where $\lambda^{2} = (E_{\rm ref}-E_{\rm pred})^{2}$ and
$U_{\alpha,\nu}$ are the derivatives of $\mathcal{L}^{NLL}$ with
respect to each variable
\begin{equation}
\left\{
\begin{array}{l}
    U_{\nu} = \frac{\beta(\nu+1)}{\alpha \nu} \\ U_{\alpha} =
    \frac{2\beta(1+\nu)}{\nu}[\exp(\Psi(\alpha+1/2)-\Psi(\alpha))-1 \\
\end{array}\right.
\end{equation}
and $\Psi(\cdot)$ is the digamma function. This model is referred to
as DER-L.  For training DER-M and DER-L, the weights for forces,
dipoles and charges were the same as for DER-S.\\

\noindent
\paragraph{Gaussian Mixtures Models}
A third alternative to quantify the uncertainty is the so-called
Gaussian Mixture Model (GMM). This method is convenient for
representing - typically - multimodal distributions in terms of a
combination of simpler distributions, such as multidimensional
Gaussians\cite{goodfellow2016deep}
\begin{equation}
    \mathcal{N}(x|\mu_{i},\Sigma_{i}) =
    \frac{1}{(2\pi)^{D/2}|\Sigma_{i}|^{1/2}}\exp{\left(-\frac{1}{2}(x-\mu_{i})^{\top}\Sigma^{-1}_{i}(x-\mu_{i})\right)}
    \label{eq:gauss}
\end{equation}
Here, $\mu_{i}$ is a $N$-dimensional mean vector and $\Sigma_{i}$ is
the $N \times N$-dimensional covariance matrix. The distribution of
data, here the distribution of molecular features, $x$, given
parameters $\theta$ can be represented as a weighted sum of
$N$-Gaussians:
\begin{equation}
    p(x|\theta) =
    \sum_{i=1}^{N}\omega_{i}\mathcal{N}(x|\mu_{i},\Sigma_{i})
    \label{eq:gm}
\end{equation}
with mixing coefficients $\omega_{i}$ obeying\cite{bishop2024deep}
$\sum_{i=1}^{N} \omega_{i} = 1$ and $0 \leq \omega_{i} \leq 1$. The
$\omega_{i}$ coefficients are the prior probability for the
$i$th-component.\\

\noindent
Following the work of Zhu \textit{et al.}\cite{zhu2023fast}, the
parameters of Equation \ref{eq:gm} ($ \theta =
\{\omega_{i},\mu_{i},\Sigma_{i}\}$) to construct the GMM were obtained from the molecular features of the last
layer of a trained PhysNet model, \textit{i.e.} one of the ensemble members.
The distribution of molecular
features from the training set is used to acquire the values of
$\theta$. The initial $\mu_{i}$ values were determined from k-means
clustering. To each Gaussian $i$ in the GMM model, a covariance matrix
$\Sigma_{i}$ is assigned. The number of Gaussian functions required
was determined by using the Bayesian Information Criterion (BIC) and
was $N=37$. Finally, the fitted model was evaluated by using the
NLL of the molecular feature vector as:
\begin{equation}
    NLL(p(x|X)) =
    -\ln{\left(\sum_{i=1}^{N}\omega_{i}\mathcal{N}(x|\mu_{i},\Sigma_{i})\right)}
\end{equation}
Here, $p(x|X)$ is the conditional probability of a molecular feature
vector $x$ with respect to the distribution of feature vectors in the
training data set $X$. The value of NLL is used as a measure of the
uncertainty prediction, whereby smaller NLL-values indicate good
agreement. The "detour" involving the feature vectors is a
disadvantage over the other methods studied here because it is not
possible to directly relate the predicted energy with the
corresponding uncertainty.\\

\subsection{Analysis}
\paragraph{Outlier detection.}
In this work, outliers are detected by considering whether a number
$N_{\rm error}$ can be found in the $N_{\rm var}$ with the highest
variance (or NLL in the case of GMM). Therefore, the accuracy for
detecting outliers is defined as:
\begin{equation}
    Acc = \frac{n(N_{\rm error} \cap N_{\rm var})}{N_{\rm var}}
    \label{eq:out_detec}
\end{equation}
Here, $n(\cdot)$ is the cardinality of the intersection between the
set of samples with the largest errors and the set with the largest
variances. Complementary to this, a classification analysis of
prediction over error and predicted variance was performed; details
can be found on the SI.\\

\paragraph{Inside-Outside distribution}
 As ML permeates more throughout daily life and is used in life-critical situations (i.e. self-driving cars\cite{nitsch2021out}, medical diagnosis\cite{zadorozhny2022out}), it is important to quantify whether identified outliers are related to a lack of information or a new discovery. As a consequence, the definition of inside-outside distribution is a controversial topic
in the ML literature. Here, the natural definition of statistical
learning theory is used:\cite{vapnik1999nature} Assume a training data
distribution $p_{\rm train}(x)$ and a testing distribution $q_{\rm
  test}(x)$; a point $x_{i}$ is defined as out-of-distribution
if\cite{farquhar2022out}
\begin{equation*}
    q_{\rm test}(x_{i}) \neq p_{\rm train}(x_{i}).
\end{equation*}

The definition described here is strict to statistical learning theory. However, other possibilities based on an energy-based criteria\cite{liu2020energy,wu2023energy}, score functions\cite{jiang2023detecting} or nearest neighbors\cite{sun2022out} can also be used. In this work, a rank is considered to assess whether
a given molecular structure is inside or outside a given distribution. First, all 28 intermolecular distances were computed. These distances were classified into
"bonded" and "non-bonded" separations as follows: if the distance is
smaller than the mean of the van der Waals radii of the two atoms
involved plus 20\%, the value is considered "bonded"; otherwise it is
non-bonded. The van der Waals radii used\cite{mantina2009consistent}
were 1.10 \AA\/, 1.70 \AA\/, and 1.52 \AA\/, for H-, C-, and
O-atoms. Next, the 28 distances were computed for all structures in
the training data set to determine $p_{\rm bond}(r)$ and $p_{\rm
  no-bond}(r)$. Using these distributions, it was possible to query a
given distance of the samples in the test data set to be inside
($Q_{5\%}(r)<r_{i}<Q_{95\%}(r)$) or outside (otherwise) the
distribution $p(r)$. Here $Q_{5\%}(r)$ and $Q_{95\%}(r)$ are the 5 \%
and 95 \% quantile of $p(r)$. Using this criterion the contribution
$\chi_{j}(r_{i})$ of distance $r_i$ for structure $j$ is
\begin{equation}
\chi_j(r_{i})=
\begin{cases}
  1  & r_{i} \in p_{\rm bond}(r) \\
  0.5 & r_{i} \in p_{\rm no-bond}(r) \\
  0 & r_{i} \notin [p_{\rm bond}(r) \cap p_{\rm no-bond}(r)]
\end{cases}
\label{eq:class}
\end{equation}
From this, $rank_j$ for sample $j$ was determined according to
\begin{equation}
    rank_j = \sum_{i}^{R} \chi_{j}(r_i)
    \label{eq:rank}
\end{equation}
where $R=28$ is the total number of distances. Using the mean of van der Waals radii to determine the
$rank$ is only one possibility. Alternative metrics based on covalent radius, bond orders or electronic densities may give different results.

\section{Results}

\subsection{Characterization of the Trained PESs}
The performance of all trained models is assessed on a hold-out test
set and the MAEs and RMSEs on energies and forces are given in
Table~\ref{sitab:maes}. While most models reach similar MAE($E)\leq
1.0$~kcal/mol, the performance on the forces deserves more attention
and is discussed further below. An essential requirement of an ML-PES
is to adequately describe geometries and relative energies of
particular structures, including the minima and transition states,
Figure \ref{fig:stat_points}. It is found that all models considered
perform adequately to predict energies of stationary points with
errors of $< 0.1$ kcal/mol. The errors for the {\it syn-}Criegee
structure are 0.01, 0.03, 0.16, -0.04, and 0.06 kcal/mol for Ens-3,
Ens-6, DER-S, DER-L, and DER-M compared with errors lower than 0.01
kcal/mol for the TS using ensembles, and -0.07,-0.01 and 0.06 kcal/mol
with DER-S, DER-L and DER-M, respectively. The smaller error of Ens-3
compared with Ens-6 is counter-intuitive and may be a consequence of
random noise in the prediction caused by, e.g., parameter
initialization, convergence of the loss function, or numerical
inaccuracies\cite{kaser2023numerical,MM.error:2024}.\\

\noindent
Complementary to the energy of the equilibrium structures, the Root
Mean Squared Displacement (RMSD) between optimized geometries from the
trained NN models and at the MP2 level were compared; see Figure
\ref{sifig:rmsd_mols}. Generally, the deviations between the obtained
geometries and the reference structures are very small. However, some
differences between the tested models can be highlighted. First, it is
noticed that models that use DER have an RMSD two or three orders of
magnitude larger than ensembles. Additionally, it is observed that the
geometry of the TS is predicted more accurately than the
\textit{(syn)}-Criegee or VHP conformations. For the DER models, the
geometries obtained with DER-S are the most accurate by approximately
two orders of magnitude compared to the ones produced with its
counterparts. On the other hand, structures obtained with DER-M have
the largest RMSD among the models tested here. The last of the DER
models tested, DER-L, produces constant RMSD for the different
molecules. Finally, the results obtained with GMM are of a slightly
lower quality than those from the ensemble models. This is expected
because the GMM model is based on one of the ensemble members. \\

\noindent
Another quantity that can be used to characterize a PES are the
harmonic frequencies for the stationary points
obtained from the Hessian matrix ($H=\partial^{2} E / \partial
\boldsymbol{r}^{2}$). The results (Figure \ref{sifig:frequencies} for {\it syn-}Criegee, TS and VHP)
indicate that the best performers are the ensemble models and GMM with
a MAE one order of magnitude lower than the DER models. Regarding the
DER models, the best performer is DER-L, followed by DER-S and
DER-M. DER-L displays errors between $-50$ cm$^{-1}$ and 50 cm$^{-1}$,
whereby most of the frequencies below 1500 cm$^{-1}$ were
underestimated and those above 2000 cm$^{-1}$ (XH stretch) were
overestimated. Conversely, DER-S underestimates most frequencies,
showing the largest errors for the vibrations at larger
frequencies. The worst performing model, DER-M, shows a large
overestimated value at around 500 cm$^{-1}$ and a large underestimated
value at high frequencies.The harmonic frequencies for the TS and for VHP follow similar
trends. It is interesting to note that the large errors in the
harmonic frequencies are also observed for the forces; in general, DER models have an MAE(F) one order of
magnitude larger than the other three models evaluated here, see
Table~\ref{sitab:maes}. This is a direct consequence and a limitation
of the assumed normal distribution of the energies. The forces and
Hessians are derivatives of the energy expression and the associated
errors are $\propto
\frac{\mathrm{Error}_{\mathrm{Ener.}}^{2}}{\sigma^{2}}$ and $\propto
\frac{\mathrm{Error}_{\mathrm{Ener.}}^{3}-\sigma^{2}}{\sigma^{4}}$,
respectively. Hence, the DER models have an inferior performance for
forces and harmonic frequencies.\\

\subsection{Calculations and Simulations with the PESs}
Next, the performance of the different PESs for reactive MD
simulations is assessed. For this, the minimum energy and minimum
dynamic paths (MEP, MDP) were computed, and finite-temperature MD simulations were carried out. The MEP describes the lowest
energy path connecting reactants and products passing through the TS. Complementary to the MEP, the
MDP\cite{unke2019sampling} provides information about the least-action
reaction path in phase space. \\

\begin{figure}
    \centering
    \includegraphics[width=0.9\textwidth]{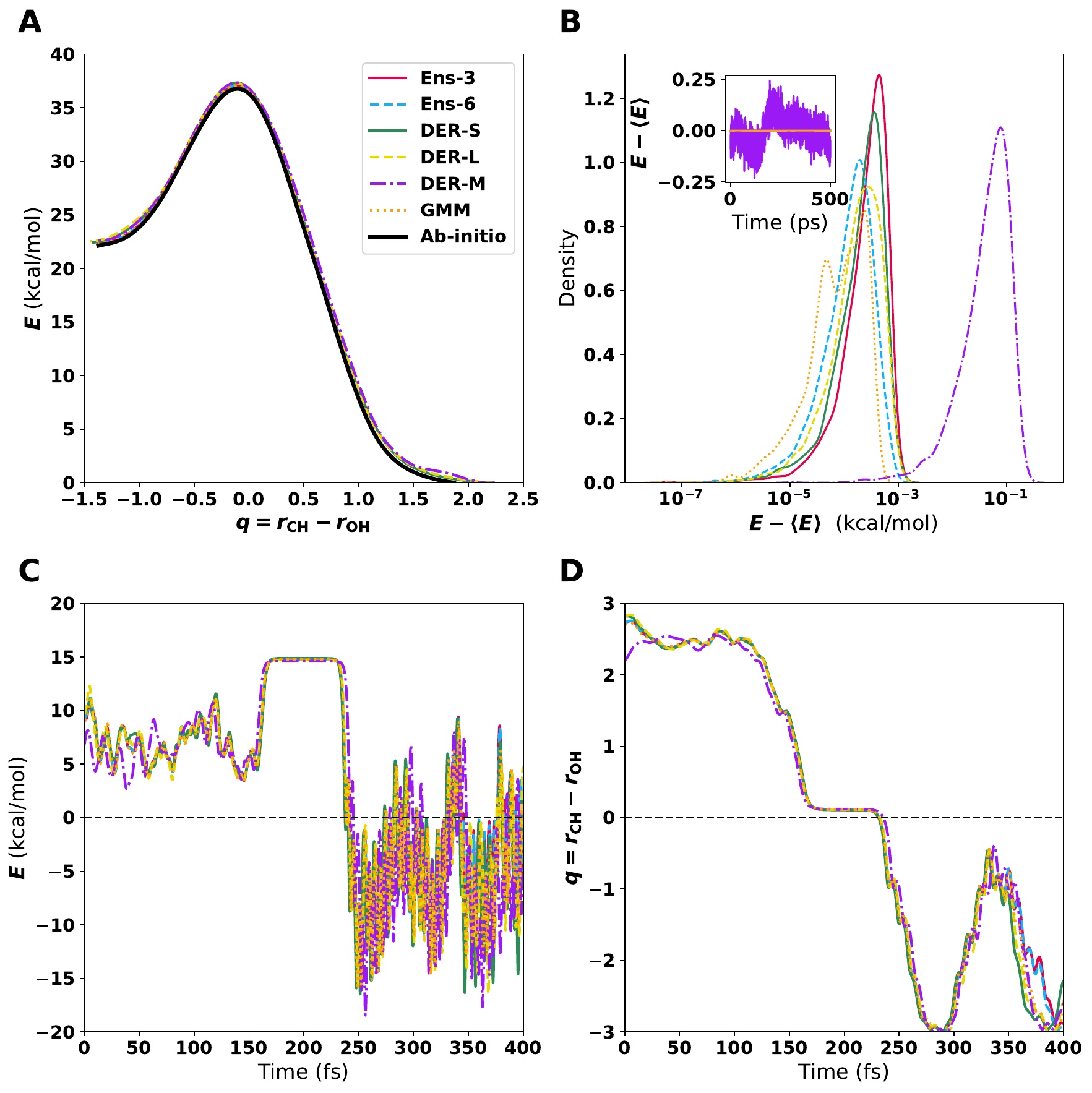}
    \caption{Behaviour of the different
      models during simulation. Panel A shows the Minimum energy path (MEP) from {\it
        syn}-Criegee to VHP for the different methods for UQ used in
      this work. The zero of energy is the corresponding value for the
      optimized structure of VHP. Panel B shows the energy distribution for the
      different models during the simulation. Note that the $x$-axis
      is on a logarithmic scale.  Starting from
      \textit{(syn)}-Criegee, the system was simulated for 500 ps with
      a time step of 0.1 fs. The inset shows the time series of the energy for DER-M. Panel C shows the variation of the energy for the
      Minimum Dynamic Path (MDP) of the different formulations of the
      ML-PESs starting from the optimized TS. Panel D
      reports the time series of the reaction coordinate ($q=r_{\rm
        CH}-r_{\rm OH}$) from the MDP. }
    \label{fig:sim_uq}
\end{figure}

\noindent
Figure \ref{fig:sim_uq}A shows the MEP for the different models
considered here. All MEPs are within less than 0.5 kcal/mol on each of
the points sampled. Therefore, despite the differences in how errors
are handled and their magnitude for each model, the MEP derived from
the PESs are consistent with one another and nearly identical. The
MDPs (see Figure \ref{fig:sim_uq}C), initiated from the TS were
determined with an excess energy of $10^{-4}$ kcal/mol.  The TS
structure is stabilized because it is a 5-membered ring and because
little excess energy was used for the MDP. VHP is observed after 225
fs accompanied by pronounced oscillations in the potential energy
primarily due to the highly excited OH-stretch. Overall, the time
traces for potential energy (Figures \ref{fig:sim_uq}C), one possible
reaction coordinate $q = r_{\rm CH} - r_{\rm OH}$ (Figures
\ref{fig:sim_uq}D), and all atom-atom separations in Figure
\ref{sifig:mdp_all_dist} are rather similar for the 6 models
considered. Notable exceptions concern primarily DER-M (purple) for
which the energy differs somewhat from the other
five models. Along similar lines, the C1-H2 and C2-H3 separations
deviate noticeably from the other 5 models; see Figure
\ref{sifig:mdp_all_dist}.  On the product (VHP) side, the
high-frequency oscillations with a period of $\sim 10$ fs (see Figure
\ref{fig:sim_uq}C) correspond to a frequency of $\sim 3500$ cm$^{-1}$
characteristic of the OH-stretch vibration, whereas the low-frequency
oscillation in Figure \ref{fig:sim_uq}D is due to the azimuthal
rotation of the -OH group. \\

\noindent
Finally, $NVE$ simulations with all six models were carried out; see
the SI for details on these simulations. The simulations were run for
500 ps with a time step of 0.1 fs, and energy is conserved to within
$\sim 0.1$~kcal/mol or better, see Figure \ref{fig:sim_uq}B. Importantly, no drift
was found on this time scale for most of the models except for
DER-M.\\

\subsection{Analysis of Error Distributions}
Next, the errors, their magnitude and distributions for the trained
models are analyzed in more detail. It is desirable that a model
accurately predicts the energies across a wide range which points
towards its extrapolation capabilities. The data set considered
contains structures for \textit{(syn)}-Criegee, VHP, and the
corresponding TS. Residual plots were used to describe
how the signed error $\Delta = E_{\rm Ref}-E_{\rm Pred}$, is
distributed for energies between $-700$ and $-300$ kcal/mol. \\

\begin{figure}
    \centering
    \includegraphics[scale=0.6]{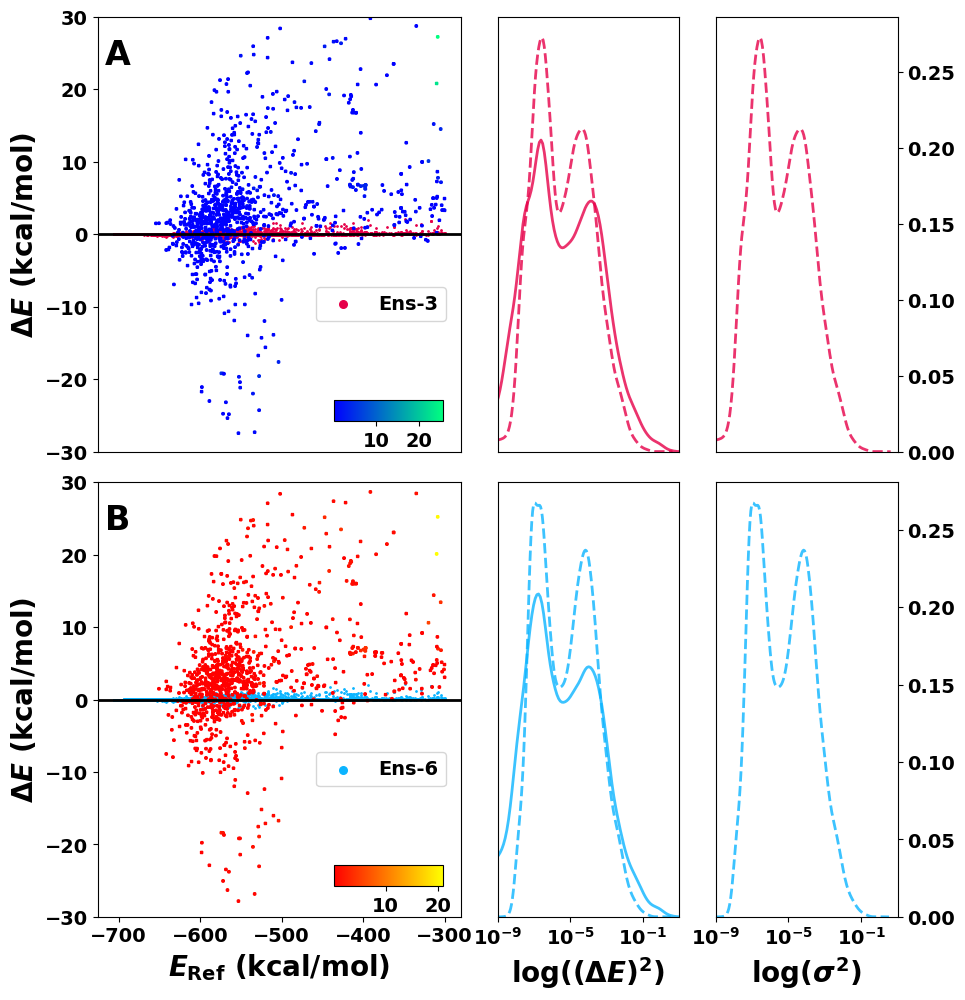}
    \caption{Performance of the Ens-3 and Ens-6 on the test
      set. Panels A and B on the left show residual plots of the error
      between reference and prediction. The 1000 energies with the
      largest variance are shaded with a different colour and directly
      reflect the model's capability to detect outliers. The
      corresponding colour bar represents the scale of the
      variance. Squared error distribution (solid lines) and variance
      distributions (dotted lines) are shown in the centre next to
      panels A and B for comparison. Complementary to this is the
      variance distribution shown on the right of both panes. Notice
      that the $x$-axis on the centre and right are in logarithmic
      scale.}
    \label{fig:ensembles_all}
\end{figure}

\paragraph{Ensembles}
Figure \ref{fig:ensembles_all} shows the performance of the ensembles.
Noticeably, the error range is between $-30$ and 30 kcal/mol, with
most errors near the centre (\textit{i.e.}  $\Delta=0$). The region
with the lowest energy ($E<-650$ kcal/mol) has higher accuracy with no
noticeable outliers. The next region, between $-650$ and $-500$
kcal/mol, have the largest number of outliers broadly spread between
positive and negative errors. For higher energies (above $-500$
kcal/mol) a small spread of the errors with few significant
outliers is found. It can be noticed that the region with more
outliers is close in energy to the transition state; therefore, the
structures are expected to have larger deformation than the other
regions. This is related to the fact that the training data set was
created to reproduce adequately the hydrogen transfer.\\

The distributions of the squared error ($P((\Delta E)^{2})$) and the
variance ($P(\sigma^2)$) in Figure~\ref{fig:ensembles_all} are both
rather sharp and centred around 0. Using a logarithmic scale further
clarifies the structure of these distributions. The bimodal nature of
$P((\Delta E)^{2})$ and $P(\sigma^2)$ is the first distinctive
feature. In addition, the predicted variance largely matches the
squared error distribution (Figure \ref{fig:ensembles_all}
centre). The distributions agree nearest to their centre. However, the
height of the distribution is larger for $P(\sigma^{2})$ than for
$P((\Delta E)^{2})$. Furthermore, the tails of $P(\sigma^2)$ decay
faster than for $P((\Delta E)^{2})$. This is reflected in fewer
samples labelled with large variance than the number of structures
with large squared error.\\

\begin{figure}
    \centering
        \includegraphics[scale=0.5]{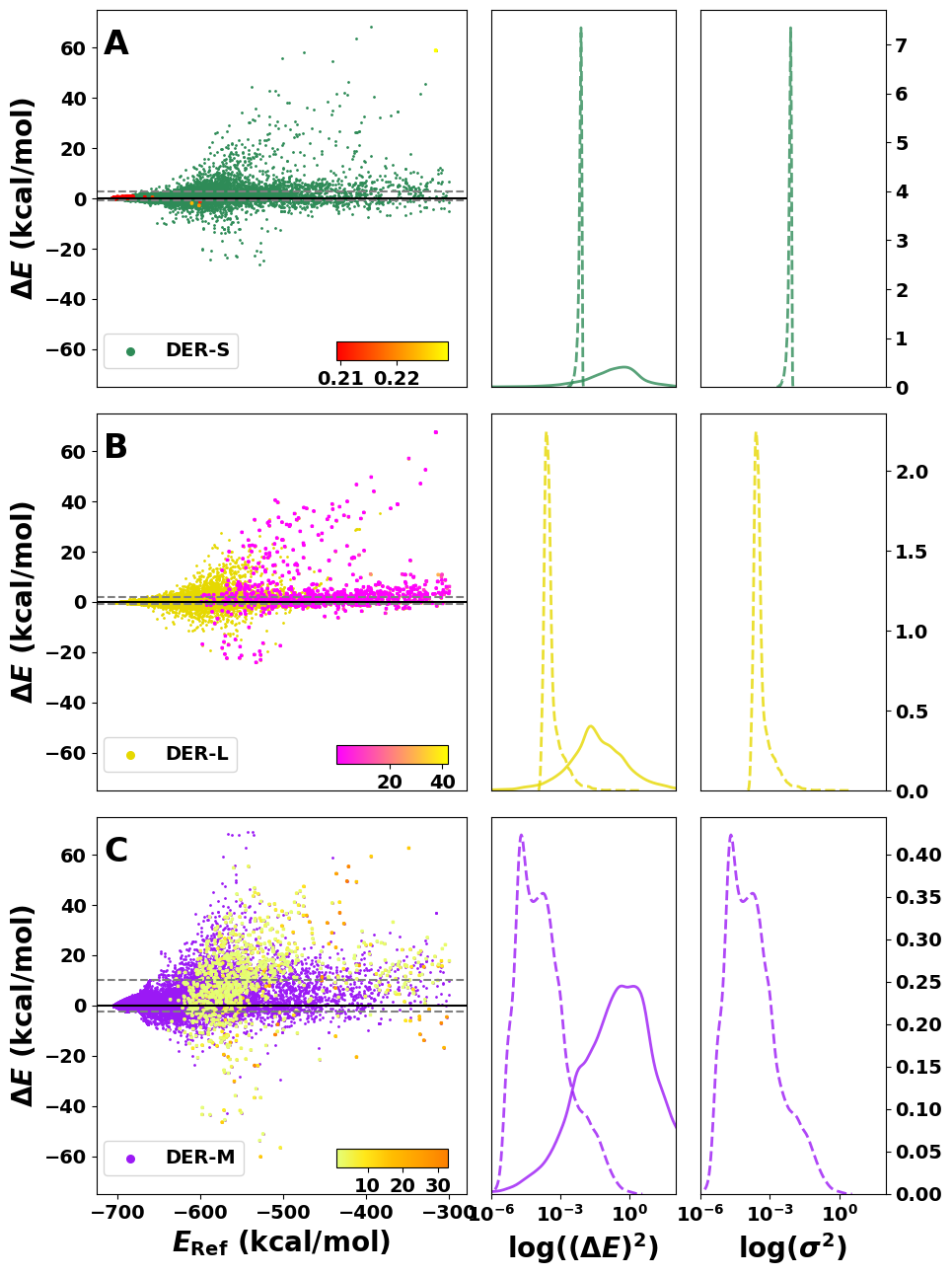}
   \caption{Performance of the different versions of PhysNet-DER
     through the range of energies of the test set. Panels A to C on
     the left show residual plots of the error between reference and
     inference for DER-S, DER-L, and DER-M, respectively. The 1000
     points with the largest variance are shaded with a different
     colour (red, magenta, and yellow from top to bottom) and directly
      reflect the model's capability to detect outliers. The
      corresponding colour bar represents the scale of
     the values.  Squared error distribution (solid lines) and
     variance distributions (dotted lines) are shown in the centre
     next to panels A, B, and C for comparison. Complementary to this
     is the variance distribution shown on the right of both
     panels. Notice that the $x$-axis on the centre and right are in
     logarithmic scale.}
    \label{fig:der-all}
\end{figure}

\paragraph{Deep Evidential Regression.}
The results for the predictions of the DER models are displayed in
Figure \ref{fig:der-all}. For DER-S, the errors are spread between
$-60$ and 60 kcal/mol, and the variances vary between $2\times10^{-3}$
to $9\times10^{-3}$ kcal/mol with a single sharp peak around $10^{-2}$
kcal/mol, \textit{i.e.} the same uncertainty for nearly all predictions. This
aligns with the previously discussed problems of
DER\cite{meinert2023unreasonable} that reported models which improve
the quality of the predictions by increasing their uncertainty. The
small variances across the test set indicate that adding forces and
dipole moments to the loss functions renders the model overconfident. One possible
explanation is that terms depending on forces, charges and
dipoles in Equation~\ref{eq:forces_der} to DER-S act as extra regularizers
to the evidence of incorrect predictions, akin to the
$\mathcal{L}^{R}(x)$ term, during training of the NN. Hence, the
variance predicted by DER-S loses its capability to detect
outliers. Furthermore, DER-S tends to underestimate the energies with
a larger population on the positive side of the $\Delta E$. Finally,
the squared error, centred around $10^{0}$, is spread over a wide
range from $10^{-4}$ to a few tens of kcal/mol.\\

\noindent
Next, DER-L is considered (see Figure \ref{fig:der-all}B) for which
the error increases with the energy. Complementary, the variance is
high for structures with positive $\Delta E$ (red points). The
variance distribution is sharply peaked and centred around $10^{-3}$,
showing some overlap with $P((\Delta E)^{2})$, whereas $P((\Delta E)^{2})$ is unimodal and centred at $10^{-1}$
kcal/mol. However, the tails are wide and extend to $10^{2}$ kcal/mol.
As for DER-S, the centre of mass of $P(\sigma^{2})$ is
between 1 or 2 orders of magnitude smaller than $P((\Delta E)^{2})$, indicating that DER-L is
overconfident about its predictions. It is also noted that DER-L is
biased to identify predictions that underestimate the energy (\textit{i.e.}, positive $\Delta E$) as outliers.\\

\noindent
Finally, DER-M (Figure \ref{fig:der-all}C) features a large dispersion
of the predicted error around the energy range considered in this
work. Predictions deteriorate quickly for low-energy configurations
with almost no points near the diagonal. $P((\Delta E)^{2})$ is centred around 1 kcal/mol and extends from $10^{-2}$
to $10^{2}$ kcal/mol with some overlap with the bimodal $P(\sigma^{2})$ centred at $\sim 10^{-4}$, around four orders of
magnitude smaller than $P((\Delta E)^{2})$. Regarding the
detection of outliers, it is found that samples which underestimate
the energy display a large variance.  On the technical side, it has
been found that optimization of multidimensional Gaussian models, such
as DER-M, can be numerically challenging because the NN-prediction of
the covariance matrices can be numerically
unstable.\cite{guo2017calibration,Seitzer2022Pitfalls,megerle2023stable}\\

\noindent
Differences between the three flavours of DER were
noticeable. Firstly, DER-M performs worst on energy predictions with a
poor quality of the underlying PES. On the other hand, DER-S and DER-L
show a similar distribution of errors; see Figure
\ref{fig:der-all}. $P(\sigma^{2})$ for DER-M is bimodal and
considerably broader than for the other two models, which show a
single sharp peak. The width of $P(\sigma^{2})$ for DER-M
increases the overlap with the $(\Delta E)^{2}$ distribution and,
therefore, is more likely to identify outliers than the other two DER
models. Unfortunately, the variance values predicted by DER-M
underestimate the error by 2 to 3 orders of magnitude. From these
results, DER-L is the best performer with the small MAE among the DER
models and medium quality for the variance estimation.\\

\begin{figure}
    \centering
    \includegraphics[width=\textwidth]{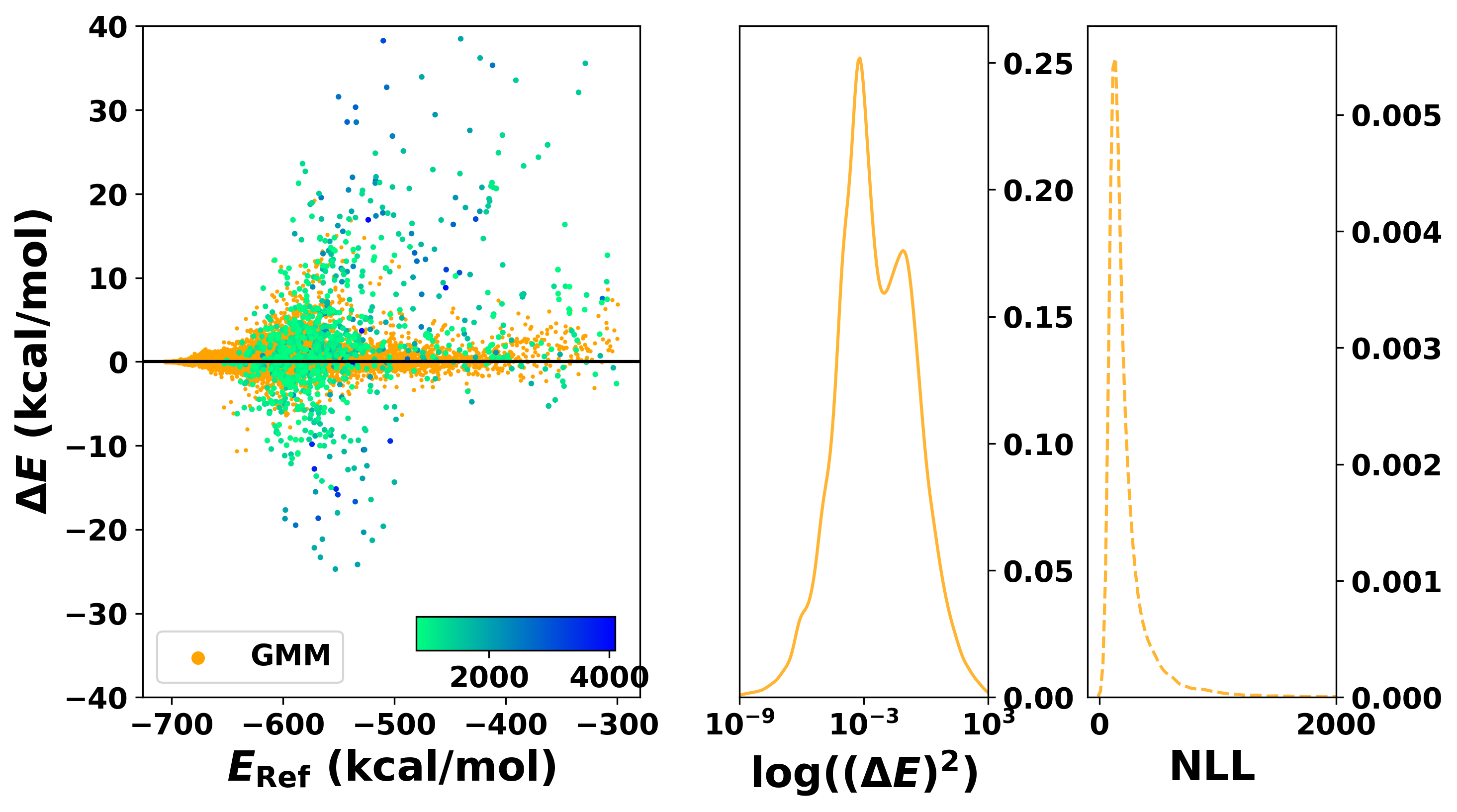}
    \caption{Performance of the PhysNet-GMM through the range of
      energies of the test set. A Residual plot of the error between
      reference and production is shown on the left. The 1000 points
      with the largest negative log-likelihood (NLL) value are shaded
      with a different colour and directly
      reflect the model's capability to detect outliers. The
      corresponding colour bar
      represents the scale of the values. The panel in the centre
      shows the squared error distribution. Note that the $x$-axis of
      the centre panel is in logarithmic scale for clarity. The panel
      on the right displays the distribution of the NLL, which is used
      to quantify the uncertainty.}
    \label{fig:gmm}
\end{figure}

\paragraph{Gaussian Mixtures Models}
Finally, for the GMM (Figure \ref{fig:gmm}), the dispersion of the
error increases as the energy increases. Specifically, the largest
errors occur for the highest energies. For the errors, it is found that
they are more evenly distributed in the over- ($\Delta E <0$) and
under-predicted ($\Delta E >0$) regions. On the other hand, $P((\Delta
E)^{2})$ features a bimodal distribution centred at $10^{-3}$ with
extended tails up to $10^{3}$ with the NLL peaked at low values of NLL
and decays rapidly for increasing NLL.\\

\subsection{Outlier Detection}
The focus of the present work is the detection of outliers. The error analysis carried out so far indicates that outlier detection is
challenging. While the high error structures are reliably captured in particular
for Ens-3, Ens-6, Der-L and GMM, they also falsely classify structures with low
errors as outliers. In this work, outlier detection capabilities of the
models are evaluated using the accuracy metric defined in Equation
\ref{eq:out_detec} and the classification procedure described in the
method section.\\

\begin{figure}[h!]
\centering
\includegraphics[width=0.9\textwidth]{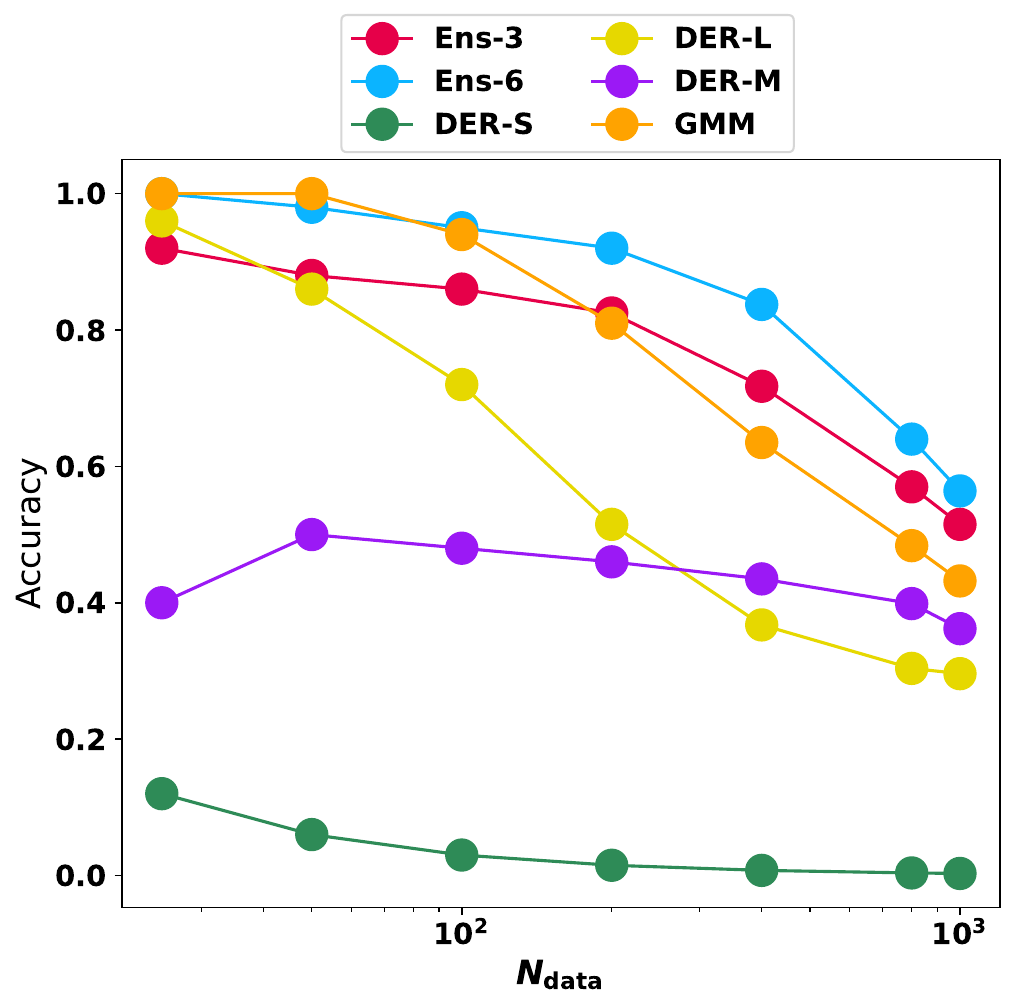}
\caption{Reliability of outlier-detection for the different
  strategies: Given the 1000 structures with the
  largest variance/uncertainty, it is evaluated whether they
  correspond to the structures that also have the largest errors from
  comparison with reference data for $N_{\rm data} = [25, 50,
    100, 200, 400, 800, 1000]$. \textit{I.e.} it is evaluated whether the $N_{\rm
    data}$ structures with the actual highest errors are contained in
  the 1000 that are predicted to have high errors.}
\label{fig:accuracy}
\end{figure}

\noindent
First, the number of structures with large variance was determined,
and the magnitude of the error was assessed. Figure \ref{fig:accuracy}
shows the results for the 1000 structures with the largest predicted
variance. The results indicate that as the number of structures with
large errors sought increases, the probability of finding them among
the top 1000 with large variance decreases. Overall, the
best-performing model is Ens-6, closely followed by Ens-3 and GMM. The
three DER models behave quite differently from one another. First,
DER-S has a poor performance and approaches zero
ability to detect outliers. Next, DER-L is very good at detecting
extreme outliers, performing even better than Ens-3 for $N_{\rm data} = 25$. However, its performance decays quickly and is the second worst 
after DER-S for $N_{\rm data} = 1000$. Finally, DER-M has an almost linear performance, meaning
its capability predictions are constant, independent of the number of
samples.\\

\noindent
One interesting aspect of Figure \ref{fig:accuracy} is that for the
extreme cases (i.e. detecting the 25 samples with the largest error),
four models (Ens-3, Ens-6, DER-L, and GMM) have a probability higher
than 80\% for detecting those extreme values. This trend continues for
the ensemble models and GMM up to $N_{\rm data} = 200$ beyond which the
accuracy decays for all models. This can be understood because the task at hand is harder to solve as the number of required samples to identify increases.  \\

\begin{figure}[h!]
\centering
\includegraphics[width=\textwidth]{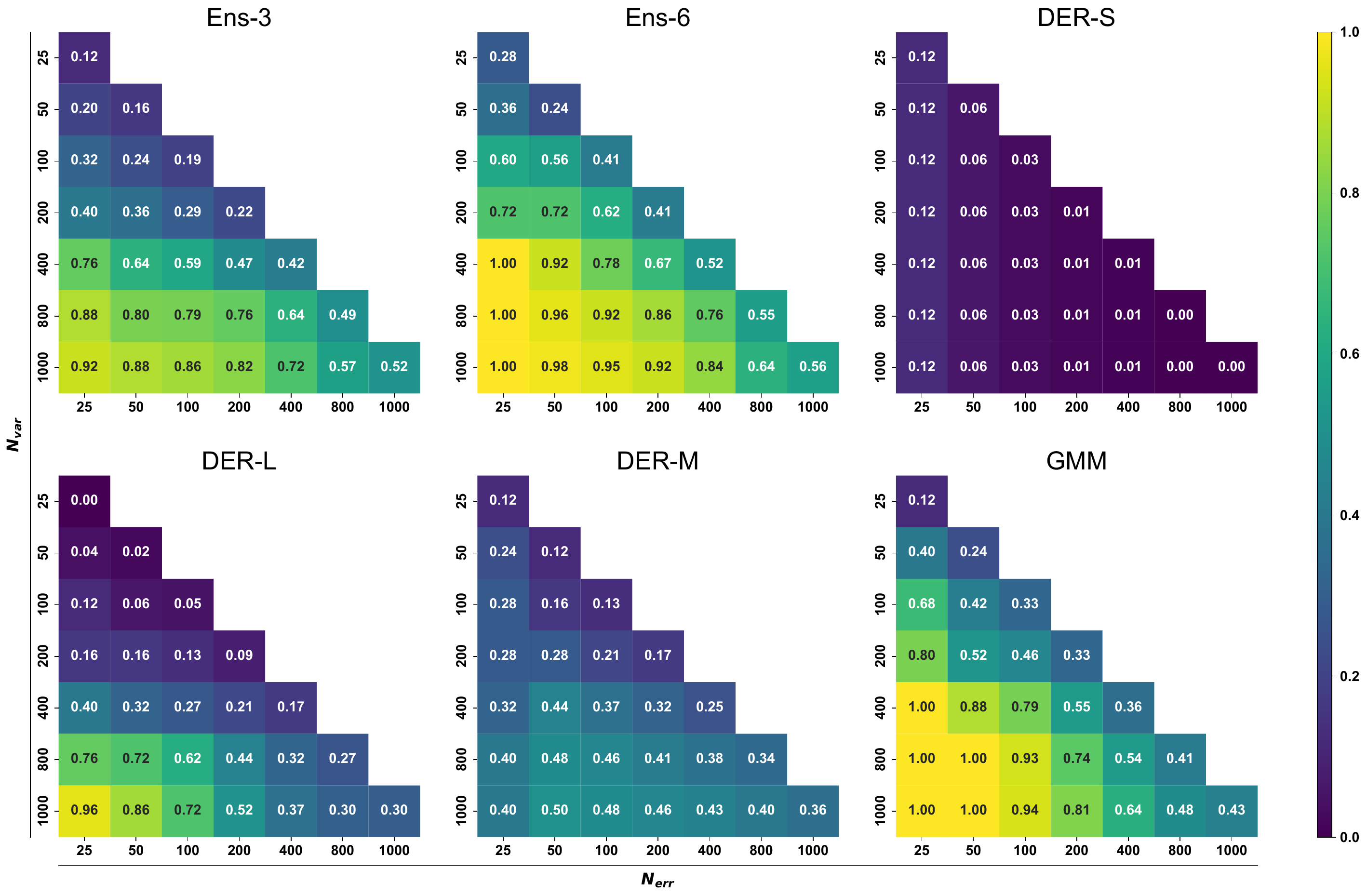}
\caption{Reliability of outlier-detection for the different
  strategies: Given $N$ structures with the highest error/variance, it
  is evaluated if they correspond to the $N$ structures with the
  largest errors/variance. See Equation \ref{eq:out_detec}. The plot
  is coloured according to the accuracy. Exact values of the accuracy
  are given for each combination in white.  }
\label{fig:accuracy_2d}
\end{figure}

\noindent
Next, a 2-dimensional analysis involving different numbers of
structures with large errors and different numbers of high-variance
structures was carried out. Figure \ref{fig:accuracy_2d} shows the
probability of finding $N_{\rm err}$ structures with large error among
the $N_{\rm var}$ structures with large variance for each
method. As an example, for Ens-3, the lower left corner reports a
probability of 0.92 for finding the $N_{\rm err} = 25$ structures with
large error among the $N_{\rm var} = 1000$ structures with large
variance. Increasing $N_{\rm err}$ to 1000 reduces this probability to
0.52. This row corresponds to the data reported in Figure
\ref{fig:accuracy}. More generally, the $N_{\rm var}$ can now be
reduced from 1000 to 25, and the probability of finding corresponding
large-error predictions is reported in the full triangle. Light and
dark colours correspond to high and low probabilities,
respectively. In practice one wants to keep $N_{\rm var}$ small and
increase the probability to find a maximum of $N_{\rm err}$
structures. From this perspective, the best-performing model is GMM.\\

\noindent
With Ens-3 as the reference, Ens-6 and GMM perform slightly better
overall, whereas DER-L is comparable for small $N_{\rm err}$ and large
$N_{\rm var}$. As $N_{\rm var}$ decreases to 400 samples and below the
reliability of DER-L drops drastically. DER-M performs inferior to
DER-L for small $N_{\rm err}$ and large $N_{\rm var}$ but maintains a
success rate of 0.2 to 0.4 for most values of $N_{\rm err}$ and
$N_{\rm var}$. Finally, DER-S has the lowest success rate throughout
except for $N_{\rm err} = N_{\rm var} = 25$ for which it performs
better than DER-L.\\

Complementary to the reliability analysis in Figures
\ref{fig:accuracy} and \ref{fig:accuracy_2d}, the true positive rate
(sensitivity or TPR, Eq. \ref{sieq:tpr}), that quantifies how many of
the samples identified with a large variance also have a large error
(\textit{c.f.} true positives), and the positive predictive values (precision
or PPV, Eq. \ref{sieq:PPV}) that measures how many of the samples with
a large error are correctly labelled by the model were analyzed. This
test was performed over different ranges of squared error and variance
(or NLL for GMM), which can be used as confidence boundaries. Ideally,
the model is expected to have large sensitivity and precision. Results
for this analysis are shown in Figures
\ref{sifig:tpr_ens3}-\ref{sifig:tpr_gmm}, which report a heatmap of
TPR and PPV values using different thresholds for error or variance in
the plot. Larger (desired) values are coloured blue while small values
are shown in red.  The results indicate that Ens-6 and Ens-3 have high sensitivity for all error ranges at low variance
values (Figure \ref{sifig:tpr_ens3} and
\ref{sifig:tpr_ens6}). Conversely, PPV values are high at all variance
ranges for a small error cutoff. It is also observed that the
confidence range for Ens-6 (Figure \ref{sifig:tpr_ens6}) is larger
than for Ens-3 (Figure \ref{sifig:tpr_ens3}). Results for the DER
models also have large TPR values at small uncertainty values (Figures
\ref{sifig:tpr_der_simple}, \ref{sifig:tpr_lips} and
\ref{sifig:tpr_MD}). On the contrary, the PPV coverage is almost null
for DER-S (Figure \ref{sifig:tpr_der_simple}) and DER-L (Figure
\ref{sifig:tpr_lips}), while DER-M has high values for all variance
ranges with a small error threshold (Figure \ref{sifig:tpr_MD}). Note, however,
that the scales for squared error and variance differ by 2 to 3 orders of magnitude. Hence, the magnitude of the MSE and MV needs to be carefully inspected in addition to the colour code. Lastly, the
TPR for GMM shows a good performance over a large range of NLL values,
which implies the model correctly assigns uncertainty to errors in a
larger range of uncertainty (Figure \ref{sifig:tpr_gmm}). On the other
hand, PPV values are obtained for large values of NLL but low squared
error threshold (Figure \ref{sifig:tpr_gmm}).\\

\noindent
Finally, two more metrics to quantify the reliability over the range
of squared errors and variance were evaluated. The first is the
false positives rate (FPR, Eq. \ref{sieq:fpr}), also known as
``false alarm rate'', which measures how many of the samples
identified with large variance do not correspond to a large error. Secondly, the false negative rate (FNR,
Eq. \ref{sieq:fnr}) or "miss rate" quantifies how many samples not
identified with a large variance correspond to a large error. For FPR
and FNR small values (red) are desirable, whereas large values (blue)
are undesirable. The results for both metrics are shown in Figures
\ref{sifig:fpr_ens3} to \ref{sifig:fpr_gmm}. For the ensemble models,
${\rm FPR} \sim 0$ over the range evaluated (Figures
\ref{sifig:fpr_ens3} and \ref{sifig:fpr_ens6}), indicating a low probability of misclassifying
samples, i.e. suitable for outlier detection. Complementary, the FNR
values are small for small variance values (Figures
\ref{sifig:fpr_ens3} and \ref{sifig:fpr_ens6} left), while the
probability of missing a sample with a large error increases with the
variance. The results for DER models show low values of FPR except for
very small values of variance (Figures
\ref{sifig:fpr_simple}, \ref{sifig:fpr_lips}, and \ref{sifig:fpr_MD}
left). Regarding the results for the FNR, large values are obtained
except for very small values of variance (Figure
\ref{sifig:fpr_simple}, \ref{sifig:fpr_lips}, and \ref{sifig:fpr_MD}
right). Finally, the GMM model has large values of FPR at low values of
NLL (Figure \ref{sifig:fpr_gmm} left) while the values of FNR are low
in a large region but decay rapidly at large values of NLL (Figure
\ref{sifig:fpr_gmm} right). These results suggest that Ens-6 is the best model for detecting outliers with high TPR, and PPV complemented with a low FPR and FNR. On the contrary, the worst model is DER-S, which has a low probability of identifying outliers.\\

\subsection{In- and Out of Distribution}
\begin{figure}
    \centering
    \includegraphics[width=0.9\textwidth,height=0.7\textheight]{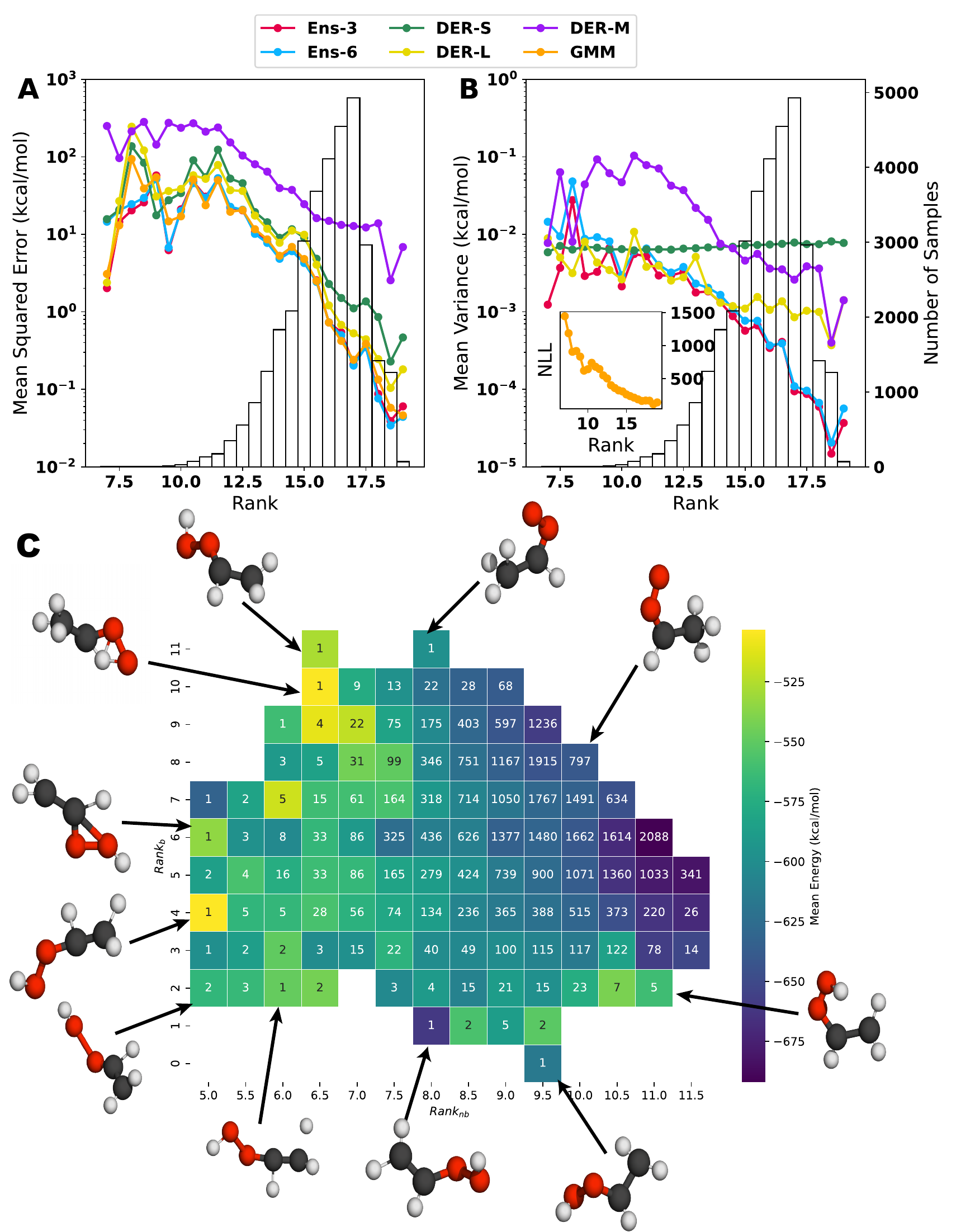}
    \caption{Evolution of the mean squared error (A) and the mean
      variance (B) concerning the rank of each structure in the test
      set. The bar plot (background) shows the number of structures
      with a particular rank. A large $rank-$value indicates that more
      degrees of freedom are covered by the training data and
      \textit{vice versa}. The $y-$axis is displayed in logarithm
      scale to highlight the difference in the values of MSE or MV for
      the different $rank$ values. Notice that for the Gaussian mixture
      model, the negative log-likelihood is used to estimate the
      uncertainty. The inset on the right panel shows how the mean
      NLL changes concerning the defined $rank$. Panel C shows the
      2d-map representation of the $rank$ for bonded and non-bonded
      separations. Representative structures of different combinations
      are shown around the map.}
    \label{fig:rankvmae}
\end{figure}

A deeper understanding of the origin of the variances and the
prediction error can be obtained by considering the distribution of
structural features (atom distances) in the training and testing
data sets, and to relate them to predicted properties. Following the procedure described in Section 2.3, a
score (the $rank$) for each molecule in the test set was
calculated. The results in Figure \ref{fig:rankvmae} are combined with
a histogram of the number of molecules with a given rank. The $rank$,
see Equations \ref{eq:class} and \ref{eq:rank}, is interpreted as the
degree to which a sample can be considered in or out of the
distribution of atom separations covered by the training set: a high $rank$ implies that more degrees of freedom (DOF)
can be found in the training data. Thus, it is  "in distribution"
(ID), while a low $rank$ indicates that the sample has more DOFs
farther away from the distribution and is "out of
distribution" (OOD). The black histogram in Figure \ref{fig:rankvmae}
shows that most samples  have $rank > 14$ and are ID to some extent, with a most probable
value $rank = 17$.\\

\noindent
Figures \ref{fig:rankvmae}A and B indicate that $rank$ and MSE or MV
(coloured lines) are related. Similarly, the distribution of samples
with given $rank$ also impacts MSE and MV, see black histograms. For
the MSE (Figure \ref{fig:rankvmae}A), all models except for DER-M
behave similarly overall. Up to $rank \sim 12$, the MSE varies
between $\sim 0$ and $\sim 100$ kcal/mol, and above the MSE decays
monotonically well below 1 kcal/mol for all models except for
DER-M. For DER-M, the behaviour is not fundamentally different, but the
magnitude of the MSE is considerably increased. The MV in Figure
\ref{fig:rankvmae}B reflects the behaviour of the MSE for DER-M, and
the same is observed for Ens-3, Ens-6, and GMM. For DER-L, the decay
of the MV with increasing rank is less pronounced, whereas for DER-S
${\rm MV} \sim 0.1$~kcal/mol throughout. One reason for the decay of MSE and MV
with increasing $rank$ is the increased number of samples for given
$rank$, $P(rank)$, see black histograms Figure
\ref{sifig:samples_v_mae}.  What distinguishes DER-M from the other
five methods is the fact that the achievable MSE remains considerably
larger for most rank-values.\\

\noindent
The relationship between $rank$ and MSE/MV can also be considered
individually for bonded and non-bonded separations; see Figure
\ref{sifig:inout_bond_nobond}. Overall, the results from Figure
\ref{fig:rankvmae}A are replicated, but the relationship between
$P(rank)$ and the MSE is yet more pronounced for bonded terms. For
small sample sizes, the MSE is large and {\it vice
  versa}. Unexpectedly, for the non-bonded separations, the behaviour
for all models except for DER-M differs: For the lowest ranks, which
are sparsely populated, the MSE increases with increasing $P(rank)$ up
to $rank = 6.5$, after which the MSE decreases monotonically. The MV,
on the other hand, behaves as expected. It is noted that for DER-S
both bonded and non-bonded separations yield an almost constant value
for the MV irrespective of $P(rank)$.\\

\noindent
The relationship between $rank$ and MEA/MV for bonded and non-bonded
separations can also be analyzed in a 2-dimensional map. First, the
average energy depending on bonded and non-bonded $rank$ is
considered; see Figure \ref{fig:rankvmae}C. This map can also be
regarded as an abstract rendering of the PES. Low-energy structures
correspond to the {\it syn-}Criegee and VHP basins, followed by
structures representative of the TS between the reactant and product
and finally, higher-lying structures dominated by larger
distortions. The majority of points (93~\%, white numbers in Figure
\ref{fig:rankvmae}C) is for $8 \leq rank_{\rm nb} \leq 11.5$ and $4
\leq rank_{\rm b} \leq 9$. These structures cover an energy range from
--700 to --300 kcal/mol with the lowest-energy structures featuring
$rank_{\rm nb} \geq 11.0$ and $rank_{\rm b} \geq 5.0$. Hence, these
are comparatively "open" structures, characteristic of an elongated
molecule such as the one considered here. Examples for such structures
are provided in Figure \ref{fig:rankvmae}C.\\

\noindent
Next, the MSE and MV are mapped onto this representation, see Figures
\ref{sifig:mae_rank} and \ref{sifig:var_rank}. Hence, the map itself
remains, but the colouration changes. For the MSE, darker colours
indicate a low error, whereas lighter colours indicate higher
errors. The regions for high MSE remain the same for all six models
considered: $5.0 \leq rank_{\rm nb} \leq 7.5$ and $2 \leq rank_{\rm b}
\leq 5$, i.e. What changes, however, is the {\it maximum} MSE which is
9 kcal/mol for Ens-3 and Ens-6 and increases up to 40 kcal/mol for
DER-M.\\

\noindent
For the MV, Ens-3 and Ens-6 are on the same scale and differ
little. The largest variances for Ens-3 and Ens-6 are observed for
similar ranks as for the MSE. On the other hand, DER-S, DER-M and
DER-L are on rather different scales ranging from $10^{-3}$ (DER-S) to
$\sim 0.1$ kcal/mol (DER-M and DER-L). DER-S returns a uniform value
for all values of $rank_{\rm b}$ and $rank_{\rm nb}$. For DER-L, the
MV is larger for $5.0 \leq rank_{\rm nb} \leq 7.5$ and $0 \leq
rank_{\rm b} \leq 9$, while DER-M displays large values for a wider
region ($rank_{\rm nb} \leq 9.5$, $rank_{\rm b}\leq 8$). Finally, the
magnitude of NLL for GMM can not be directly compared with the other
five models, but NLL is large for $rank_{\rm nb} \leq 8$, $rank_{\rm
  b}\leq 8$. \\

\noindent
The preceding analysis showed that a simple ranking such as the one presented here
can highlight the effect of the differences between training and test distribution on
the prediction and the uncertainty estimation. It must be mentioned
that the $rank-$metric can be used as a proxy for how structure and
error are related. However, further analysis is required to complement
these results because averaging effects can play an important
role. Yet, for improving reactive ML-PESs it is notable that samples
with larger $rank$ feature lower average error and {\it vice
    versa}. It is also found that coverage of the non-bonded distances for predicting energies and uncertainties can be rather informative. This contrasts with the usual focus on sufficiently covering the range of chemical bonds when conceiving data sets for training ML-PESs.\\

\section{Discussion and Conclusions}
The present work analyzed quantitatively to what extent
three different UQ-methods - ensembles, Deep Evidential Regression,
and Gaussian Mixture Models - are capable to detect outliers in
samples from which full-dimensional reactive potential energy surfaces
can be trained. The system investigated for this was one of the CIs
{\it syn-}Criegee, CH$_3$CHOO.\\

\begin{figure}
    \centering \includegraphics[scale=0.8]{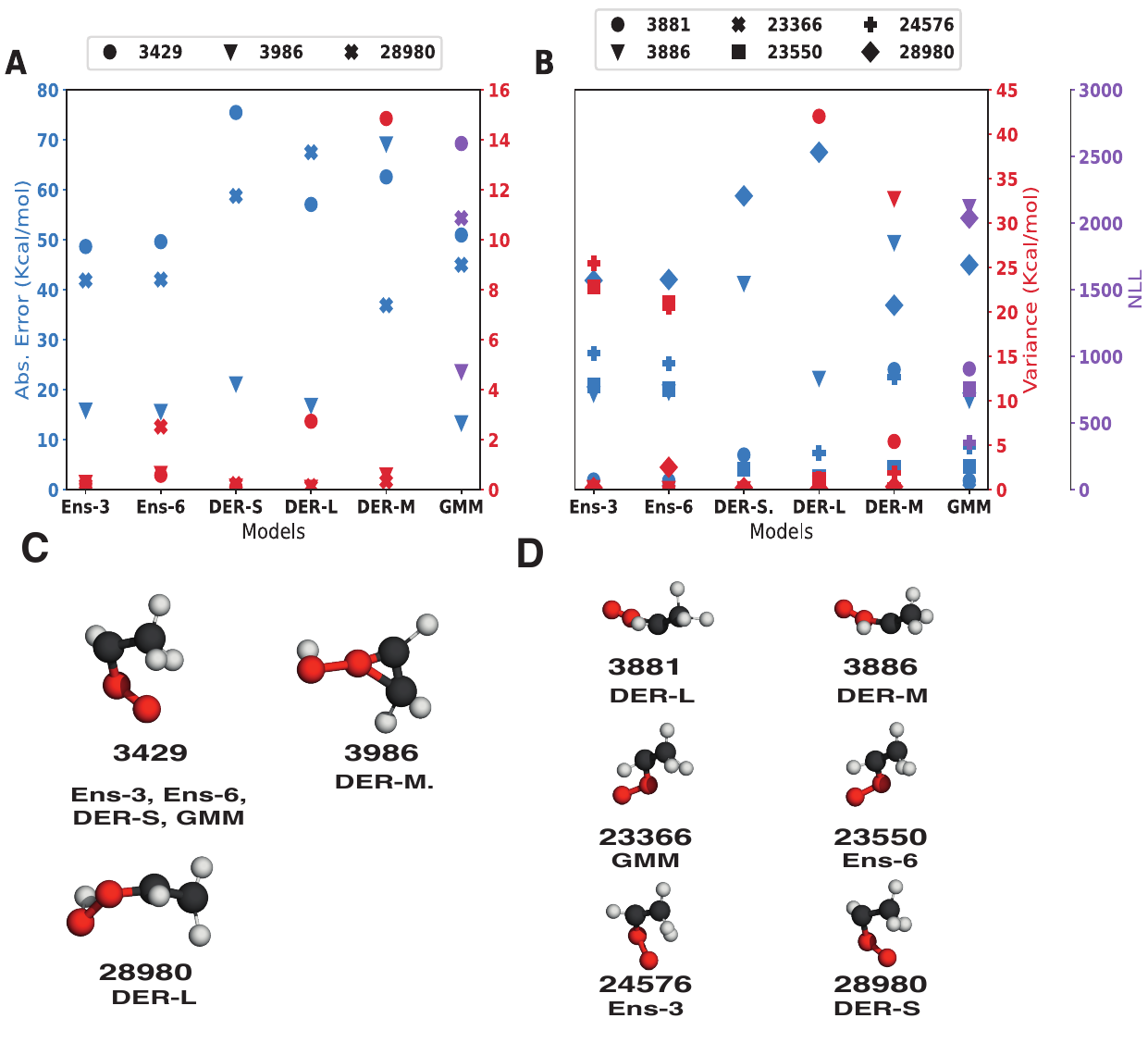}
    \caption{Extreme values in prediction. Panel A shows the values of
      the absolute error (blue) and variance (red) or NLL (purple) for
      each of the samples identified to have the largest error and its
      corresponding index. Molecular structures are shown in panel C
      with their corresponding index and the model for which the
      structure is identified to have the largest error. Panel B is
      similar to panel A but for the structures identified to have the
      largest variance. The corresponding structures are shown in
      panel D.}
    \label{fig:max_vals}
\end{figure}

From an electronic structure perspective, CIs are known to be
challenging because they feature multi-reference (MR)
effects.\cite{dawes:2015,MM.criegee:2023} This can also be
demonstrated from the present data and even be linked to the quality
of the prediction and the MV. For this, molecular structures with the
largest absolute errors (Figure \ref{fig:max_vals}A) and with the
largest uncertainty (Figure \ref{fig:max_vals}B) for each of the
models were determined. Generally, the largest errors arise either for
deformed \textit{(syn)}-Criegee or VHP structures, whereas structures
with the largest variance are predominantly perturbed
\textit{(syn)}-Criegee structures except for GMM, which identifies one
structure closer to the TS. Interestingly, none of the models assigns
the largest uncertainty to the structure with the largest error. In
all cases, the magnitude of the error is larger than the predicted
variance. On the other hand, for structures with large variance, the
errors are on the same scale for ensembles and DER-M, whereas they are
almost constant for DER-S. Contrary to this, DER-L overestimates the
uncertainty by one order of magnitude.\\

\noindent
Structure \#3429 (see Figure \ref{fig:max_vals}C) with the
largest error is the same for four out of the six models. The
remaining two models also show a large error for this structure,
indicating that this structure is, in general, difficult to
predict. Surprisingly, structure \#3429 is predicted to have a large
uncertainty for the models that do not identify it with the largest
error (DER-M and DER-L), while the other four identify it with 
smaller uncertainty. Structure \#3986 is most difficult to predict
with DER-M, while for the other models, it is better predicted with a
difference between predictions of $\approx 50$ kcal/mol. The GMM model
assigns it a large uncertainty while the other models give it values
in the same range as the predicted structure \#3986. Lastly, structure
\#28980 features the largest error for DER-L but in the same magnitude
as the other models except for DER-M. Regarding the uncertainty, Ens-6
identifies \#28980 with a large uncertainty, while the other models
attribute a small value to it. It is also found that Ens-3, Ens-6,
DER-S, and GMM identify structures (e.g. \#23366, \#23550, \#24576,
\#28980) that resemble those with the largest error; however, the
error for these four structures is not large; see SI for a
discussion.\\

\noindent
One possible reason for the difficulties in predicting energies for
particular geometrical arrangements concerns the MR
character of its electronic structure. To prove this, the
$T_{1}$\cite{lee1989diagnostic} and $D_{1}$\cite{janssen1998new}
diagnostic coefficients were determined, see Table
\ref{sitab:mr_error}. All structures with large errors clearly display
MR character which are not captured from the
single-reference MP2 reference data used in the present
work. Interestingly, the uncertainty prediction of the models appears
to be related to the MR effects as well (Table \ref{sitab:mr_var})
because the molecules identified with large variance also have large
values of $T_{1}$ and $D_{1}$ diagnostic. These findings are also
consistent with earlier work on acetaldehyde.\cite{MM.acet:2020}\\

From the present analysis, ensemble models emerge as a viable route
for outlier detection. The capability of the modified DER models are
considerably improved over DER-S, which is largely unsuitable for this
task. On the other hand, DER-L is able to detect extreme cases with
almost the same quality as the ensemble models thanks to the
modifications of the loss function (\textit{c.f.} Equation
\ref{eq:lf_lipz}). However, this capability decays rapidly with the
number of required samples $N_{\rm err}$. Finally, DER-M has a
constant probability of detecting outliers regardless of the number of
samples considered. This is an interesting behaviour because it
implies a strong correlation between the error in prediction and the
variance. Unfortunately, the probability of detecting outliers for
DER-M is $\sim 40$ \% throughout. The last model, GMM, showed an
intermediate performance between ensembles and DER. However, the NLL
as the uncertainty measure is only qualitative and can not be used
directly to estimate the error. Nevertheless, it performed well in
detecting outliers with good reliability that decay at the same rate
as ensemble models. \\

The fundamental insights
  gained from the present work are as follows. It is possible to carry
  out meaningful outlier detection for reactive PESs with the
  most successful approaches reaching 50 \% detection quality for a
  pool of 1000 structures with the highest uncertainty. Two new
  formulations of the deep evidential regression method, DER-M and
  DER-L, were presented and evaluated. The most promising among the
  approaches tested here are ensemble methods and DER-L, and it is
  found that Ens-6 and GMM yield consistent results overall. Large values of the $rank$ metric (a geometry-based descriptor) were found to correlate with large average errors suggesting that rapid-to-evaluate geometrical criteria may be an efficient way to detect outliers. These could subsequently be used to complement a given training set. A related structure-based procedure was successfully used for choosing structures best suited for transfer learning PESs for a specific process.\cite{mm.tlrpimalonaldehyde:2022} Potential future
  developments and improvements concern additional modifications to
  the loss function (scaled-by-variance\cite{Seitzer2022Pitfalls}, \textit{post-hoc} recalibration of the uncertainty using isotonic regression\cite{kuleshov2018accurate}) and exploring methods independent on the underlying statistics (such as Gaussian distribution of the data in DER) including conformal prediction
  methods.\cite{angelopoulos2021gentle,hu2022robust}\\

\section*{Supporting Information Available}
The supporting information provides further details on the methods (derivation of equations for DER-L, set up of the neural network training, classification procedure, setup of energy conservation simulations, and determination of MR character), complementary discussion, tables, and figures. 

\section*{Data and Code Availability Statement}
The code and data supporting this research will be available at \url{https://github.com/MMunibas/outlier} when the manuscript is accepted.

\section*{Acknowledgment}
This work was supported by the Swiss National Science Foundation
through grants $200020\_219779$ and $200021\_215088$ and the
University of Basel.

\bibliography{references.bib}

\appendix

\renewcommand{\thepage}{S\arabic{page}}
\renewcommand{\thetable}{S\arabic{table}}
\renewcommand{\thefigure}{S\arabic{figure}}
\renewcommand{\theequation}{S\arabic{equation}}
\setcounter{page}{1}
\setcounter{figure}{0}
\section{Supplementary methods}

\subsection{Details for DER Multidimensional}
For DER-M the outputs are constructed to be part of the covariance
matrix $\mathbf{L}$ defined as
\begin{equation*}
    (\mathbf{L})_{ij} = \left\{
    \begin{array}{ll}
        \mathrm{SoftPlus}(\ell_{i}) + \epsilon& {\rm If} \ i=j  \\
         l_{ij} + \epsilon & {\rm If} \ i>j \\
         0 & {\rm else}
    \end{array}
    \right.
\end{equation*}
Here, $\ell_{ij}$ are the outputs of the last layer ($E_{\rm
  pred},Q_{\rm pred}$) of the modified PhysNet model. It must be
mentioned that $\mathbf{L}$ is a lower triangular matrix. A difference
between the original formulation of Meinert and
Lavin\cite{meinert2021multivariate} and the one presented here is that
the exponential function for the covariance matrix is replaced with
the SoftPlus activation. Additionally, $\epsilon=1\times10^{-6}$ is
added to each of the outputs of the last layer as a
  regularizer. These modifications avoid numerical instabilities
and/or singularities during training. \\

\noindent
The parameter $\nu$ corresponds to the number of degrees of freedom of
the distribution\cite{murphy2023probabilistic}, and it is also an
output of the PhysNet model. Meinert and
Lavin\cite{meinert2021multivariate} relate $\nu$ to the number of
virtual measurements of the variance. The value of $\nu$ is
constrained to $\nu\in[3,13]$; the lower boundary corresponds to the
requirement that $\nu >n + 1$, where $n$ is the number of predicted
quantities. The upper boundary is $\nu < 13$ because it is empirically
known that for $\nu \geq 13$ the resulting distribution is
indistinguishable from a normal
distribution\cite{brereton2015t}. Then, the expression for $\nu$ is:
\begin{equation*}
    \nu = 10\left(\frac{\tanh(x)+1}{2}\right) + 3
\end{equation*}

\noindent
The aleatoric (data) and epistemic (knowledge) uncertainty of the
multidimensional model are obtained from
\begin{equation}
    \mathbb{E}[\sigma^{2}]=\frac{\nu}{\nu-3}\mathbf{L}
    \mathbf{L}^{\top}
    \label{sieq:alea_unc}
\end{equation}
\begin{equation}
    Var[\mu] = \frac{\mathbb{E}[\sigma^{2}]}{\nu}
    \label{sieq:epis_unc}
\end{equation}

\subsection{Set up of the NN training}
The neural network model used in this work is PhysNet\cite{MM.physnet:2019}. The original version in tensorflow was used for the ensemble method, while the Pytorch version was employed for DER. Five modules were used in both cases, each with two residual atomic modules and three residual interaction modules. The output of it was pooled into one residual output model. The number of radial basis functions was kept at 64,
and the dimensionality of the feature space was 128. A batch size of 32 and a learning rate of 0.001 were
used for training. An exponential learning rate scheduler with a decay factor of
0.1 every 1000 steps and the ADAM optimizer\cite{kingma2014adam} with
a weight decay of 0.1 were employed. An exponential moving average for
all the parameters was used to prevent overfitting. A validation step
was performed every five epochs.

\subsection{Classification}
Following the
methodology presented by Kahle and Zipoli\cite{Kahle2022}, we
classified the predictions obtained by the different models to
determine if the predicted uncertainty can be used as a reliable
estimation of the prediction error. In this case, the following
classes were defined:
\begin{itemize}
    \item True Positive (TP): $\varepsilon_{i}>\varepsilon^{*}$ and
      $\sigma_{i}>\sigma^{*}$.
      
    \item False Positive (FP): $\varepsilon_{i}<\varepsilon^{*}$ and
      $\sigma_{i}>\sigma^{*}$
      
    \item True Negative (TN): $\varepsilon_{i}<\varepsilon^{*}$ and
      $\sigma_{i}<\sigma^{*}$. 
      
    \item False Negative (FN): $\varepsilon_{i}>\varepsilon^{*}$ and
      $\sigma_{i}<\sigma^{*}$. 
\end{itemize}
As a difference from our previous
approach\cite{vazquez2022uncertainty}, we report the results when
$\varepsilon^{*} = {\rm MSE}$ (mean squared error) and
$\sigma^{*}=\mathrm{MV}$ (mean-variance) and also for different values
of $\sigma^{*}$ and $\varepsilon^{*}$ to obtain decision boundaries
for the relationship between variance and error.  For the different
values of $\sigma^{*}$ and $\varepsilon^{*}$, common metrics of the
overall performance were evaluated. In this work, we use the true
positive rate ($R_{\mathrm{TP}}$) or {\it sensitivity}. This quantity
is defined as\cite{watt2020machine}:
\begin{equation}
    R_{\mathrm{TP}} = \dfrac{N_{\mathrm{TP}}}{N_{\mathrm{TP}}+N_{\mathrm{FN}}}
    \label{sieq:tpr}
\end{equation}
Here, $N_{\mathrm{TP}}$ refers to the number of true positives and
$N_{\mathrm{FN}}$ to the number of false negative samples. A large
sensitivity value indicates that the model is unlikely to relate large
variance values with small errors (c.f. false negatives). \\

\noindent
Complementary to Equation \ref{sieq:tpr} is the positive predictive
value ($P_{\mathrm{TP}}$) or {\it precision}:
\begin{equation}
    P_{\mathrm{TP}} = \dfrac{N_{\mathrm{TP}}}{N_{\mathrm{TP}}+N_{\mathrm{FP}}}
    \label{sieq:PPV}
\end{equation} 
where all the previous quantities keep their meaning and
$N_{\mathrm{FP}}$ is the number of false positives. This quantity
relates to how many of the samples predicted with high uncertainty
correspond to a large error. \\

In addition it is desirable to quantify how often the model
misclassifies a prediction. This can be measured by the False Positive
Rate (FPR), which measures how many samples are classified with large
uncertainty but low error. This is defined as:

\begin{equation}
    R_{\mathrm{FP}} = \frac{N_{\mathrm{FP}}}{N_{\mathrm{FP}} +N_{\mathrm{TN}}}
    \label{sieq:fpr}
\end{equation}

\noindent
The opposite case can be quantified by the False Negative Rate (FNR) defined as:

\begin{equation}
    R_{\mathrm{FN}} = \frac{N_{\mathrm{FN}}}{N_{\mathrm{TP}} +N_{\mathrm{FN}}}
    \label{sieq:fnr}
\end{equation}

\subsection{Further Analysis of Structures / Outliers}
Structures identified with a large variance present more considerable
variations for error and variance than the corresponding structures
with the biggest values of error; in the following, we will describe
the error and variance for each structure following the enumeration of
the samples. Test structure \#3881 is related to the largest
uncertainty for DER-L; this is only replicated by DER-M and GMM, which
also assigns it a large uncertainty. However, the energy prediction is
accurate for most of the models evaluated except for DER-S. Next,
structure \#3886 has the largest uncertainty for DER-M, while none of
the other models associates it with a large uncertainty
value. Nevertheless, this structure is hard to predict for all of
them, with errors between 50 and 20 kcal/mol. Continuing with our
analysis, molecule \#11467 is discussed. This sample is identified
with the largest uncertainty value for the GMM model. Nonetheless, all
models, even GMM, perform well in predicting this sample. Structures
\#23550 and \#24576 are identified with the largest variance for the
models Ens-6 and Ens-3, respectively. Both structures are similar,
with a difference in the orientation of the carbon atom attached to
the O-O in the \textit{(syn)}-Criegee complex. Both samples show
problems to be predicted by the ensemble models; however, it looks
like models based on DER show fewer difficulties. Regarding the
predicted uncertainty for \#23550 and \#24576, GMM assigns it a large
uncertainty while the DER models assign it a low uncertainty.  Last
but not least is sample \#28980, which is identified with the largest
variance for the DER-S model. This sample is hard to predict for all
models, being the hardest for DER-L, which yields the largest error
for it. Regarding the uncertainty, it is noticed that for most of the
models, with the exception of Ens-6, the predicted uncertainty is
low. This analysis clearly shows that the prediction error is
comparable for most of the analyzed models. However, detecting this
error is not easy, as none of the extreme uncertainty values predicted
are related to the extreme error. \\

\subsection{Evaluating the Multi-Reference Character of a Structure}
Determining if a single reference method adequately describes a
molecular system is challenging. Therefore, several diagnostic metrics
have been proposed to evaluate multireference effects on the
system. Among them is the $T_{1}$
diagnostic\cite{lee1989theoretical,lee1989diagnostic}, which is the
Euclidean norm of the single substitution amplitudes vector ($t_{1}$)
of the closed-shell Couple-Cluster Single Doubles (CCSD) wave function
divided by the square root of the number of correlated electrons:
\begin{equation}
    T_{1} = \frac{\norm{t_{1}}}{\sqrt{N_{\rm corr. elec.}}}
\end{equation}
A single reference method will perform correctly if the value of the
$T_{1}$ diagnostic\cite{wang2015multireference} is
$T_{1}<0.02$. Complementary, the $D_{1}$ diagnostic
\cite{janssen1998new} is defined as the maximum Euclidean norm of the
vectors formed by the product of the matrix $\boldsymbol{S}$ which
elements $s_{1}^{2}$ are the single excitation amplitudes of the CCSD
wavefunction. Then, the $D_{i}$ diagnostic is defined as:
\begin{equation}
    D_{1}=||\boldsymbol{S}||_{2} = \max_{||x||_{2}=1}{\left(||\boldsymbol{S}\Vec{x}||_{2}\right)}
\end{equation}
Here $\boldsymbol{S}\in\mathbb{R}^{o\times v}$ with $o$ and $v$
denoting the number of active occupied and active virtual orbitals. For $D_{1} > 0.05$ the molecule is dominated by dynamic
correlation\cite{wang2015multireference}. $T_{1}$ and $D_{1}$ are
suggested to be used together because $T_{1}$ represents an average
value for the complete molecule, which might fail to indicate problems
for small regions of the molecule. In those cases, $D_{1}$ can be used
as evidence if the molecule has regions that single reference methods
can not adequately describe. \\

In this work, $T_{1}$ and $D_{1}$ diagnostics were determined for the
structures identified with the largest error for each model tested and
those with the largest uncertainty value. Then, each molecule was
computed at the CCSD(T)-F12 level of theory with the aug-cc-pVTZ basis
function with the MOLPRO suite\cite{MOLPRO}. Then, the values of
$T_{1}$ and $D_{1}$ are reported on Table \ref{sitab:mr_error} for the
molecules with large errors and Table \ref{sitab:mr_var} for those
with large variance.

\subsection{Energy Conservation Simulations}
The energy conservation of the models was estimated by running
molecular dynamics simulations over the generated potentials using the
Atomic Simulation Environment (ASE)\cite{larsen2017atomic}. $NVE$
simulations were run using Verlet dynamics. The initial velocities
were assigned to follow a Maxwell-Boltzmann distribution at 300 K. The
simulation was run from the \textit{(syn)}-Criegee intermediate for
0.5 ns using a time step of 0.1 fs. The energies were saved for every
1000 steps.

\clearpage
\section{Supplementary Tables}

\begin{table}[]
\begin{tabular}{l|cc|cc}
Model      & MAE($E$) & RMSE($E$) & MAE($F$) & RMSE($F$) \\ \hline \hline
Ens-3      & 0.44   & 1.80     & 1.54   & 11.98   \\
Ens-6      & 0.43   & 1.79    & 1.48   & 11.47   \\
DER-S & 1.03   & 2.61    & 32.06  & 90.60    \\
DER-L   & 0.69   & 2.35    & 31.79  & 90.09   \\
DER-M     & 2.19   & 5.17    & 33.55  & 91.54   \\
GMM        & 0.47       & 1.83        & 1.68       & 9.73 \\ \hline       
\end{tabular}
\caption{Summary of the statistical metrics of the predictions of energy and forces for the models tested in this work. The first two columns correspond to the values for energies, while the last two columns are the values for forces. Units are kcal/mol for energies and (kcal/mol)$\cdot \rm \AA^{-1}$ for forces.  }
\label{sitab:maes}
\end{table}

\newpage
\begin{table}[]
\begin{tabular}{lrrrrrrr}
\textbf{s-Cri.} & \textbf{MP2 Ref.}&\textbf{Ens-3} & \textbf{Ens-6} & \textbf{DER-S} & \textbf{DER-L} & \textbf{DER-M} & \textbf{GMM} \\\hline \hline
1  & 224.2   & 225.9   & 222.8   & 251.7   & 170.8   & 218.2   & 223.8   \\
2  & 304.0   & 298.8   & 297.2   & 337.0   & 273.3   & 479.0   & 300.0   \\
3  & 481.5   & 476.2   & 475.8   & 440.0   & 460.6   & 518.8   & 475.7   \\
4  & 698.5   & 691.6   & 691.2   & 679.2   & 687.9   & 686.6   & 691.2   \\
5  & 745.3   & 738.2   & 738.1   & 710.9   & 761.6   & 750.6   & 738.8   \\
6  & 939.6   & 928.3   & 928.6   & 919.0   & 924.2   & 951.9   & 927.9   \\
7  & 996.4   & 998.7   & 998.3   & 1000.7 & 993.0   & 1010.8 & 998.8   \\
8  & 1031.1 & 1035.1 & 1034.8 & 1018.6 & 1019.7 & 1067.4 & 1035.3 \\
9  & 1130.3 & 1132.2 & 1132.0 & 1112.8 & 1118.6 & 1237.4 & 1132.4 \\
10 & 1295.6 & 1286.9 & 1287.4 & 1305.4 & 1300.1 & 1328.2 & 1288.6 \\
11 & 1397.6 & 1397.4 & 1397.4 & 1379.0 & 1390.9 & 1387.2 & 1397.1 \\
12 & 1456.6 & 1451.3 & 1451.2 & 1403.4 & 1450.5 & 1441.0 & 1451.1 \\
13 & 1474.2 & 1471.3 & 1471.2 & 1484.4 & 1486.9 & 1494.1 & 1471.2 \\
14 & 1514.3 & 1513.1 & 1513.5 & 1541.3 & 1525.9 & 1540.5 & 1514.6 \\
15 & 3047.8 & 3044.2 & 3045.1 & 3060.4 & 3030.3 & 2818.8 & 3046.7 \\
16 & 3101.5 & 3088.9 & 3090.2 & 3148.6 & 3069.4 & 3085.0 & 3091.0 \\
17 & 3207.3 & 3206.2 & 3206.5 & 3171.7 & 3198.7 & 3126.4 & 3210.1 \\
18 & 3253.2 & 3255.9 & 3255.9 & 3186.7 & 3301.3 & 3253.7 & 3256.9 \\\hline
\textbf{MAE} & - & 4.7 & 4.5 & 27.3 & 17.9 & 46.5 & 4.4\\\hline
\end{tabular}
\caption{Harmonic frequencies of \textit{(syn)}-Criegee: \textit{Ab initio}
MP2 reference values are compared to the frequencies determined on the different PESs.} 
\label{sitab:freq_cr}
\end{table}

\newpage
\begin{table}[]
\begin{tabular}{lrrrrrrr}
\textbf{TS} & \textbf{MP2 Ref.}&\textbf{Ens-3} & \textbf{Ens-6} & \textbf{DER-S} & \textbf{DER-L} & \textbf{DER-M} & \textbf{GMM} \\\hline \hline
1  & 518.0   & 517.4  & 517.4  & 494.4  & 506.0  & 453.4  & 517.4  \\
2  & 533.0   & 528.5  & 528.5  & 541.3  & 524.2  & 502.2  & 528.4  \\
3  & 745.3   & 744.9  & 744.9  & 715.7  & 721.0  & 686.0  & 744.9  \\
4  & 770.9   & 768.6  & 768.6  & 765.7  & 748.2  & 766.5  & 768.6  \\
5  & 857.7   & 853.7  & 853.7  & 845.5  & 846.0  & 833.7  & 853.8  \\
6  & 969.9   & 964.0  & 964.0  & 929.0  & 932.1  & 973.2  & 964.0  \\
7  & 1010.3  & 1007.4 & 1007.4 & 992.2  & 1000.4 & 1011.2 & 1007.4 \\
8  & 1036.7  & 1033.2 & 1033.2 & 1030.7 & 1042.4 & 1063.8 & 1033.3 \\
9  & 1223.2  & 1221.3 & 1221.3 & 1201.0 & 1220.8 & 1184.8 & 1221.3 \\
10 & 1281.6  & 1281.2 & 1281.2 & 1272.0 & 1296.0 & 1250.9 & 1281.2 \\
11 & 1360.3  & 1360.0 & 1360.1 & 1329.5 & 1382.1 & 1412.2 & 1360.0 \\
12 & 1504.5  & 1503.3 & 1503.3 & 1466.9 & 1510.7 & 1555.6 & 1503.2 \\
13 & 1557.9  & 1554.2 & 1554.2 & 1542.5 & 1564.2 & 1572.7 & 1554.2 \\
14 & 1875.3  & 1866.3 & 1866.4 & 1795.1 & 1805.2 & 2021.0 & 1866.1 \\
15 & 3116.3  & 3118.9 & 3118.8 & 3095.8 & 3071.0 & 3124.2 & 3118.7 \\
16 & 3237.2  & 3236.3 & 3236.3 & 3215.1 & 3130.6 & 3235.1 & 3236.0 \\
17 & 3251.9  & 3252.9 & 3252.9 & 3230.5 & 3159.4 & 3264.3 & 3252.8 \\
$i$ & 1523.0  & 1518.3 & 1518.2 & 1574.3 & 1544.7 & 1331.7 & 1518.5 \\ \hline
\textbf{MAE} & - & 2.8 & 2.8 & 25.3 & 28.9 & 42.3 & 2.8\\\hline
\end{tabular}
\caption{Harmonic frequencies of transition state: \textit{Ab initio}
MP2 reference values are compared to the frequencies determined on the different PESs.} 
\label{sitab:freq_ts}
\end{table}

\newpage
\begin{table}[]
\begin{tabular}{lrrrrrrr}
\textbf{VHP} & \textbf{MP2 Ref.}&\textbf{Ens-3} & \textbf{Ens-6} & \textbf{DER-S} & \textbf{DER-L} & \textbf{DER-M} & \textbf{GMM} \\\hline \hline
1  & 149.1   & 178.3  & 178.7  & 194.1  & 176.2  & 209.8  & 178.7  \\
2  & 253.1   & 254.4  & 254.4  & 258    & 240.6  & 250.9  & 254.5  \\
3  & 332.5   & 331.8  & 331.8  & 338.5  & 338.5  & 376.4  & 331.8  \\
4  & 612.4   & 613.0  & 613.0  & 622.4  & 595.6  & 562.0  & 613.2  \\
5  & 711.2   & 708.7  & 708.7  & 668.5  & 626.0  & 602.0  & 708.8  \\
6  & 843.8   & 840.7  & 840.6  & 797.6  & 783.9  & 796.6  & 840.8  \\
7  & 878.3   & 876.1  & 876.0  & 878.5  & 859.2  & 839.3  & 876.2  \\
8  & 972.2   & 968.2  & 968.3  & 909.0   & 878.1  & 890.1  & 968.4  \\
9  & 975.0   & 971.7  & 971.7  & 994.6  & 988.4  & 1030.6 & 971.9  \\
10 & 1158.8  & 1156.3 & 1156.2 & 1152.8 & 1153.6 & 1130.2 & 1156.4 \\
11 & 1319.1  & 1319.1 & 1319.1 & 1340.6 & 1372.8 & 1270.9 & 1319.1 \\
12 & 1374.2  & 1372.6 & 1372.6 & 1350.4 & 1388.3 & 1325.9 & 1372.7 \\
13 & 1428.7  & 1425.4 & 1425.4 & 1449.6 & 1417.4 & 1464.7 & 1425.4 \\
14 & 1693.6  & 1691.6 & 1691.6 & 1704.9 & 1711.8 & 1689.2 & 1691.6 \\
15 & 3216.3  & 3222.5 & 3222.4 & 3144.9 & 3178.9 & 3191.5 & 3222.3 \\
16 & 3236.0  & 3235.3 & 3235.2 & 3178.8 & 3237.1 & 3299.5 & 3235.1 \\
17 & 3330.0  & 3333.7 & 3333.8 & 3313.4 & 3289.2 & 3393.3 & 3333.4 \\
18 & 3762.9  & 3759.0 & 3758.9 & 3716.0 & 3765.8 & 3821.4 & 3758.8 \\ \hline
\textbf{MAE} & - & 3.9 & 4.0 & 28.5 & 28.8 & 48.1 & 3.9 \\\hline
\end{tabular}
\caption{Harmonic frequencies of VHP: \textit{Ab initio}
MP2 reference values are compared to the frequencies determined on the different PESs.} 
\label{sitab:freq_vhp}
\end{table}

\newpage
\begin{table}[]
    \centering
    \begin{tabular}{c|c|c}
        Molecule & $T_{1}$ & $D_{1}$  \\ \hline \hline
         3429 & 0.09 & 0.45 \\
         3986 & 0.05 & 0.23 \\
         28980 & 0.05 & 0.24 \\ \hline
    \end{tabular}
    \caption{Diagnostics for assessing the multireference character of the structures identified with the largest error in the test dataset. These quantities are unitless. A value of $T_{1}>0.02$ indicates a multireference character, and $D_{1} > 0.05$ points to dynamical multireference effects.\cite{wang2015multireference}}
    \label{sitab:mr_error}
\end{table}

\begin{table}[]
    \centering
    \begin{tabular}{c|c|c}
        Molecule & $T_{1}$ & $D_{1}$  \\ \hline \hline
         3881 & 0.07 & 0.25 \\
         3886 & 0.08 & 0.35\\
         23366 & 0.04 & 0.19 \\
         23550 & 0.05 & 0.24 \\
         24576 & 0.07 & 0.36 \\ 
         \hline
    \end{tabular}
    \caption{Diagnostic metrics for the multireference character of the structures identified with the largest uncertainty in the test dataset. A value of $T_{1}>0.02$ indicates a multireference character. Complementary, $D_{1} > 0.05$ indicates the presence of dynamical multireference effects.}
    \label{sitab:mr_var}
\end{table}

\clearpage
\section{Supplementary Figures}
\begin{figure}[h!]
\centering
\includegraphics[width=0.9\textwidth]{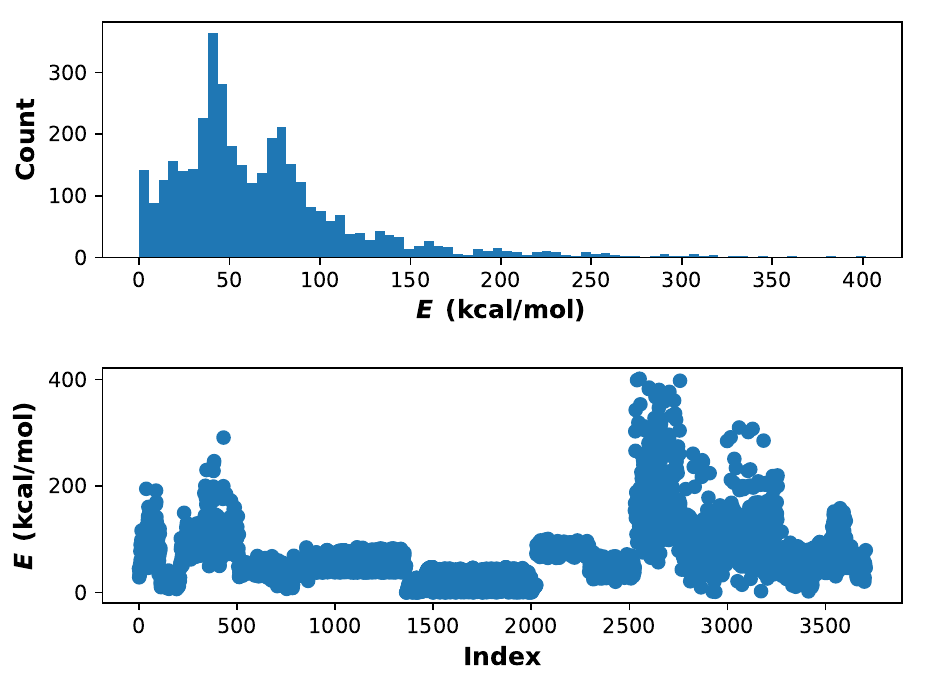}
\caption{Energy distribution of the data set employed to train the first generation ML-PES. }
\label{sifig:gen1_data_energydistr}
\end{figure}

\begin{figure}[h!]
\centering
\includegraphics[width=0.9\textwidth]{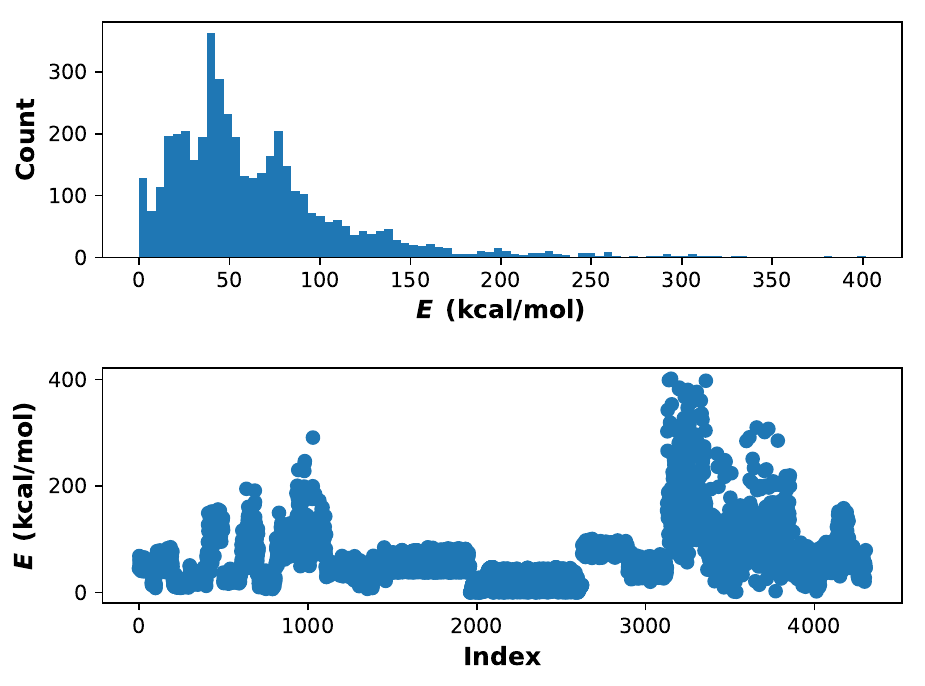}
\caption{Energy distribution of the data set employed to train the final generation ML-PES. }
\label{sifig:gen3_data_energydistr}
\end{figure}

\begin{figure}[h!]
\centering
\includegraphics[width=0.9\textwidth]{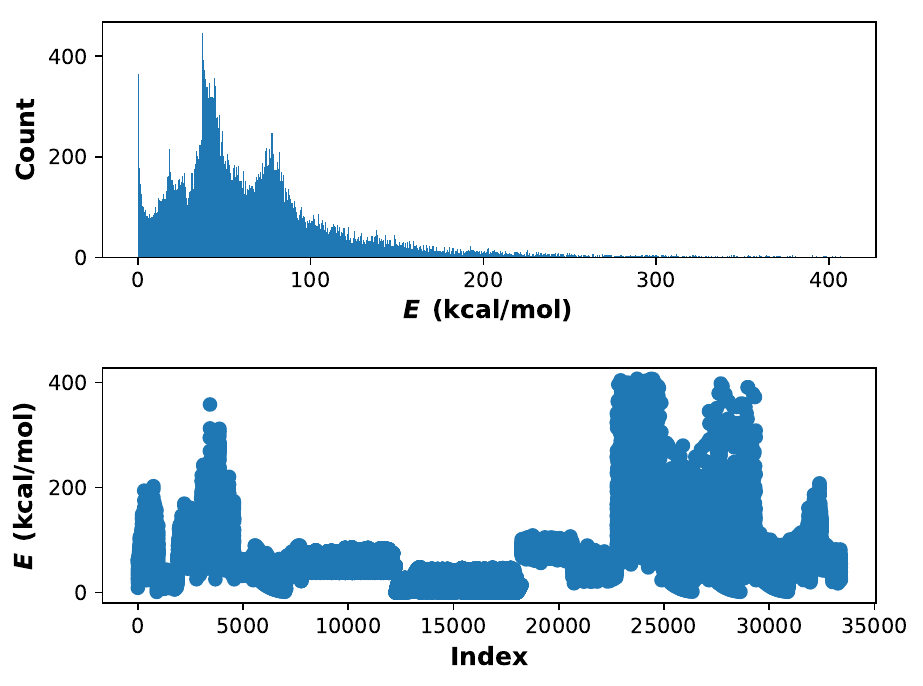}
\caption{Energy distribution of the test data set.}
\label{sifig:remaining_data_energydistr}
\end{figure}

\clearpage

\begin{figure}
    \centering
    \includegraphics[scale=0.45]{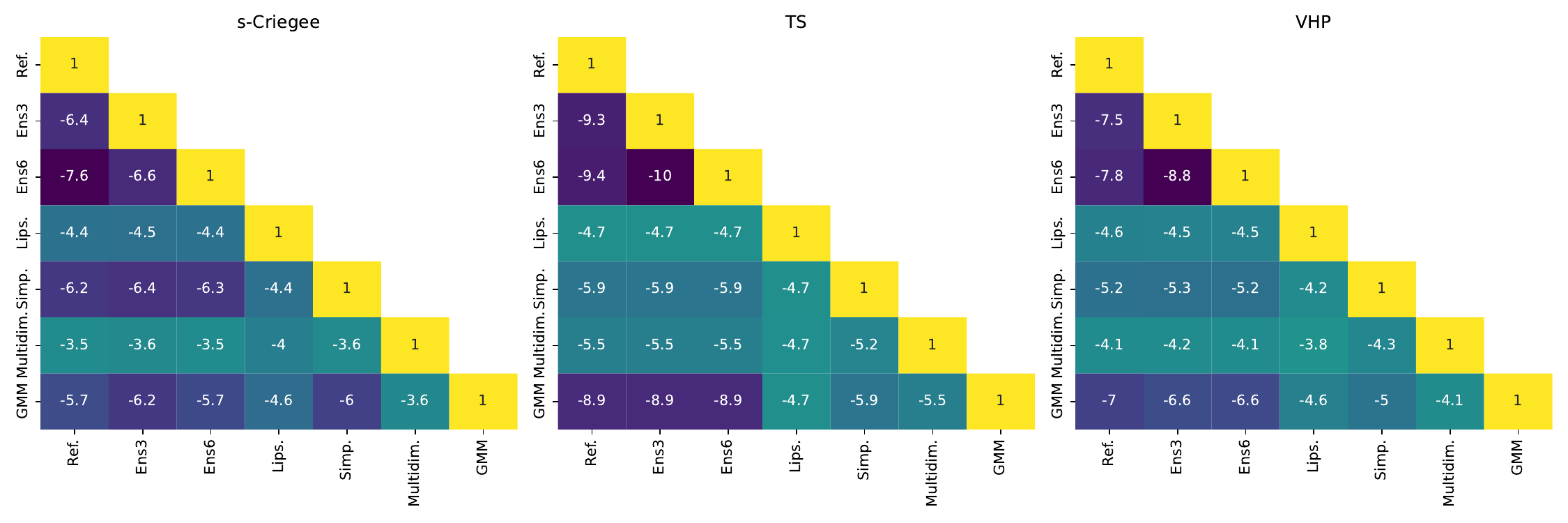}
    \caption{Root Mean Square Displacement of the stationary points (VHP, Transition State and S-Criegee) of the potential energy surface with respect to the \textit{ab-initio} reference structure and between the different obtained geometries. Notice that the logarithm of the value of RMSD is reported to exemplify the differences between the values better.}
    \label{sifig:rmsd_mols}
\end{figure}

\begin{figure}
    \centering
    \includegraphics[scale=0.7]{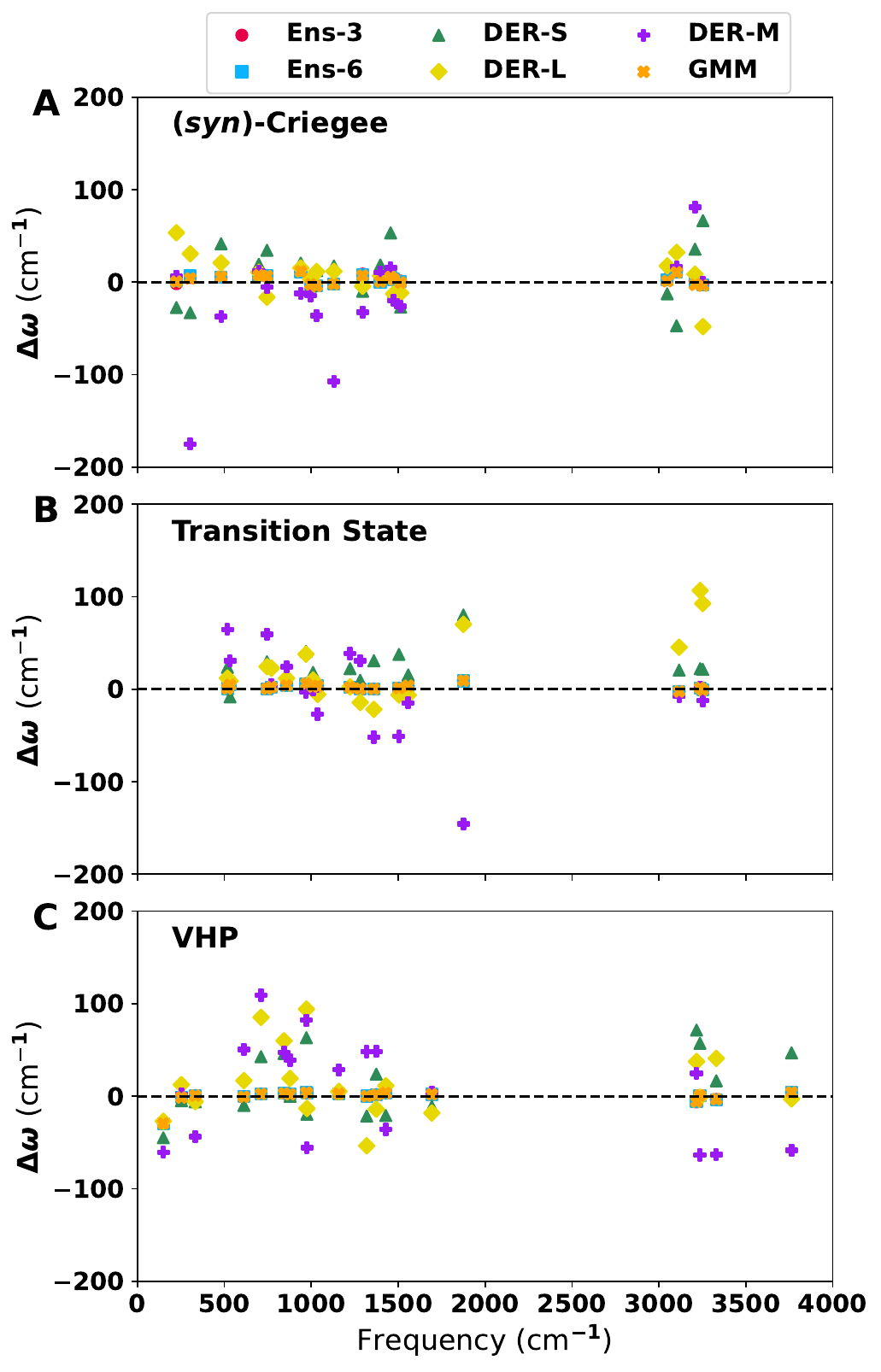}
    \caption{Error per predicted harmonic frequency ($\Delta\omega = \omega_{\rm ref}-\omega_{\rm pred}$) of the \textit{(syn)}-Criegee (A), transition state (B) and VHP (C) for all the UQ methods evaluated in this work. The values of the frequencies are reported in Tables \ref{sitab:freq_cr}, \ref{sitab:freq_ts}, and \ref{sitab:freq_vhp}. }
    \label{sifig:frequencies}
\end{figure}

\begin{figure}
    \centering
    \includegraphics[width=\textwidth]{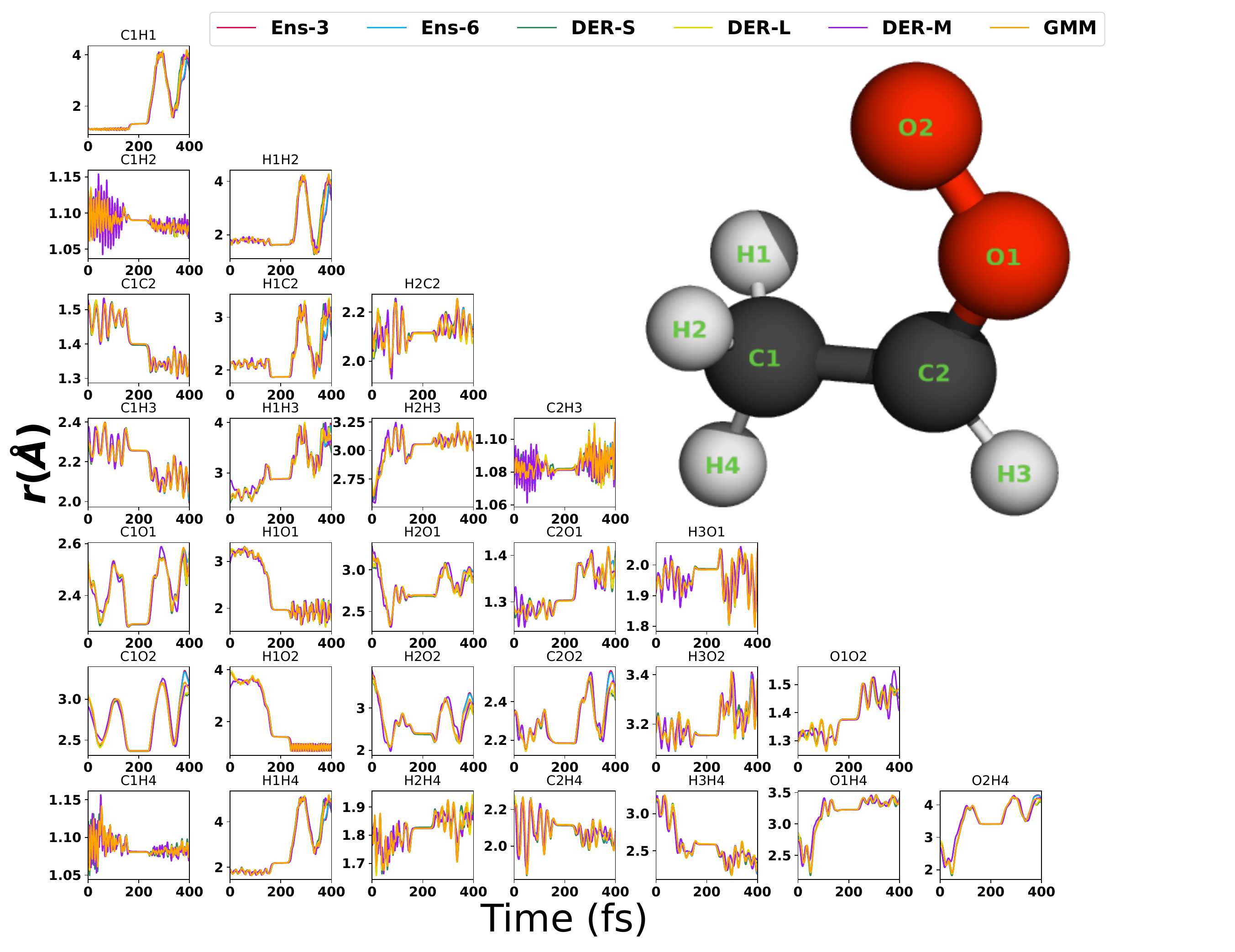}
    \caption{Atom-atom separation time series along the MDP for all models tested in this work. Each panel reports the distance between two atoms. The inset molecule displays the labelling of the atoms.}
    \label{sifig:mdp_all_dist}
\end{figure}

\begin{figure}
    \centering
    \includegraphics[width=0.9\textwidth]{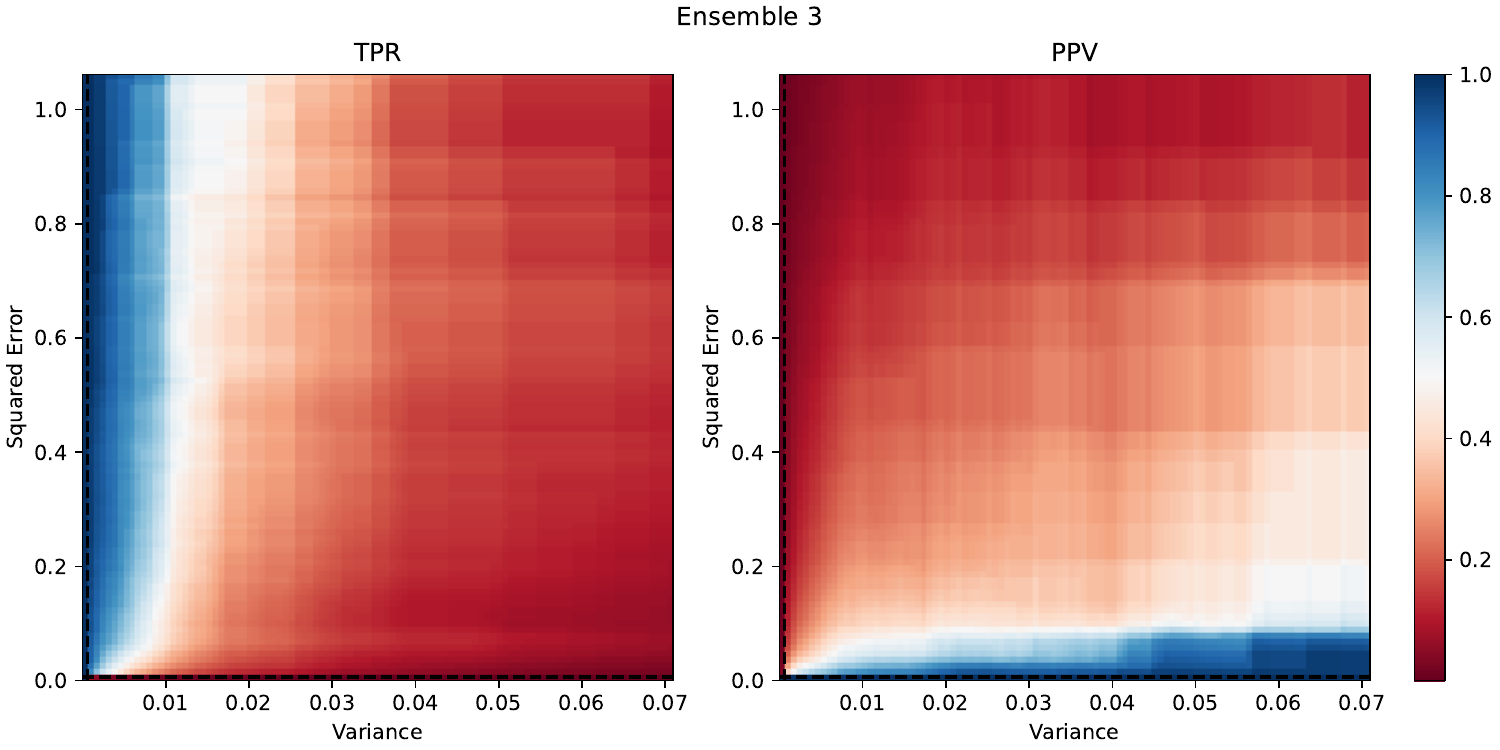}
    \caption{True Positive Rate (Left) and Positive Predictive Value (Right) for the Ens-3 model.}
    \label{sifig:tpr_ens3}
\end{figure}

\begin{figure}
    \centering
    \includegraphics[width=0.9\textwidth]{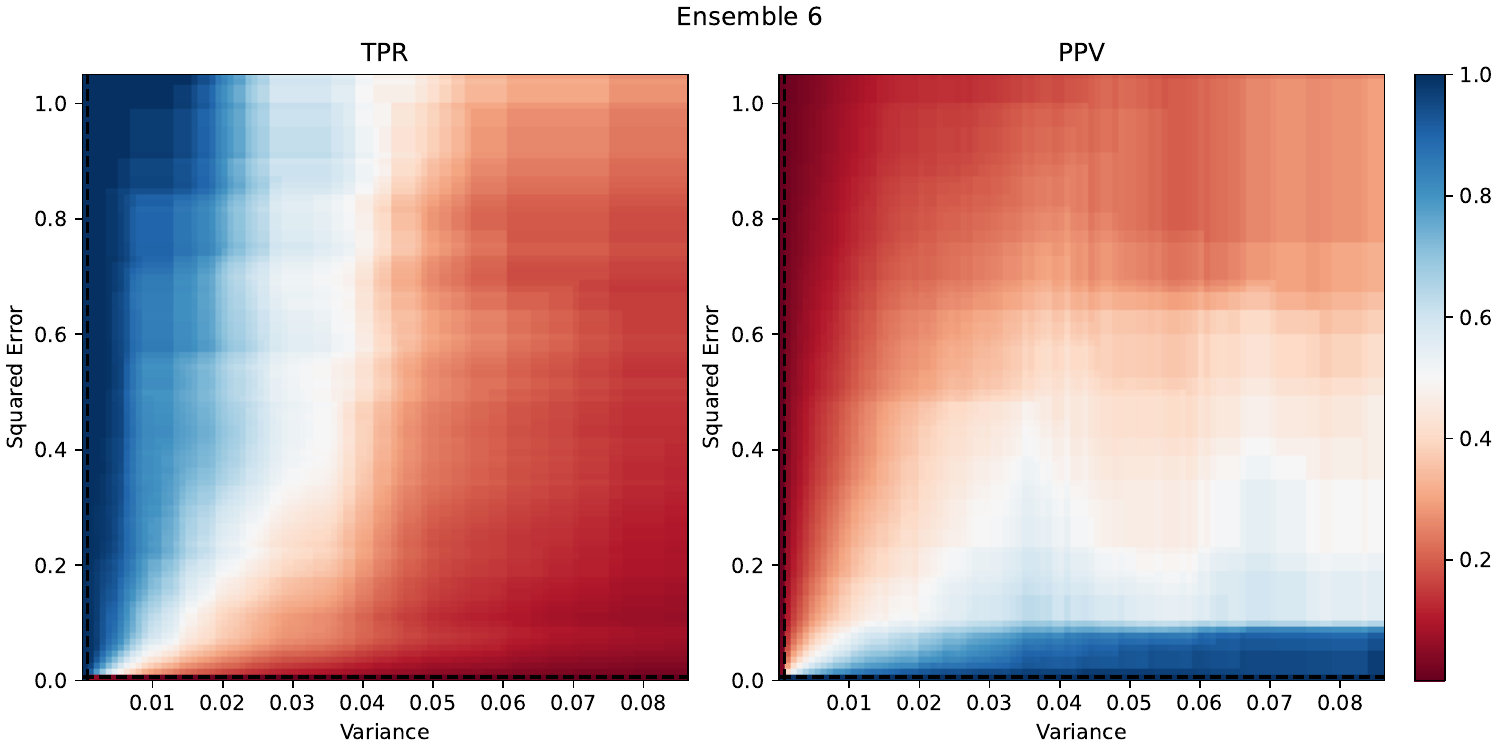}
    \caption{True Positive Rate (left) and Positive Predictive Value (right) for the Ens-6 model.}
    \label{sifig:tpr_ens6}
\end{figure}

\begin{figure}
    \centering
    \includegraphics[width=0.9\textwidth]{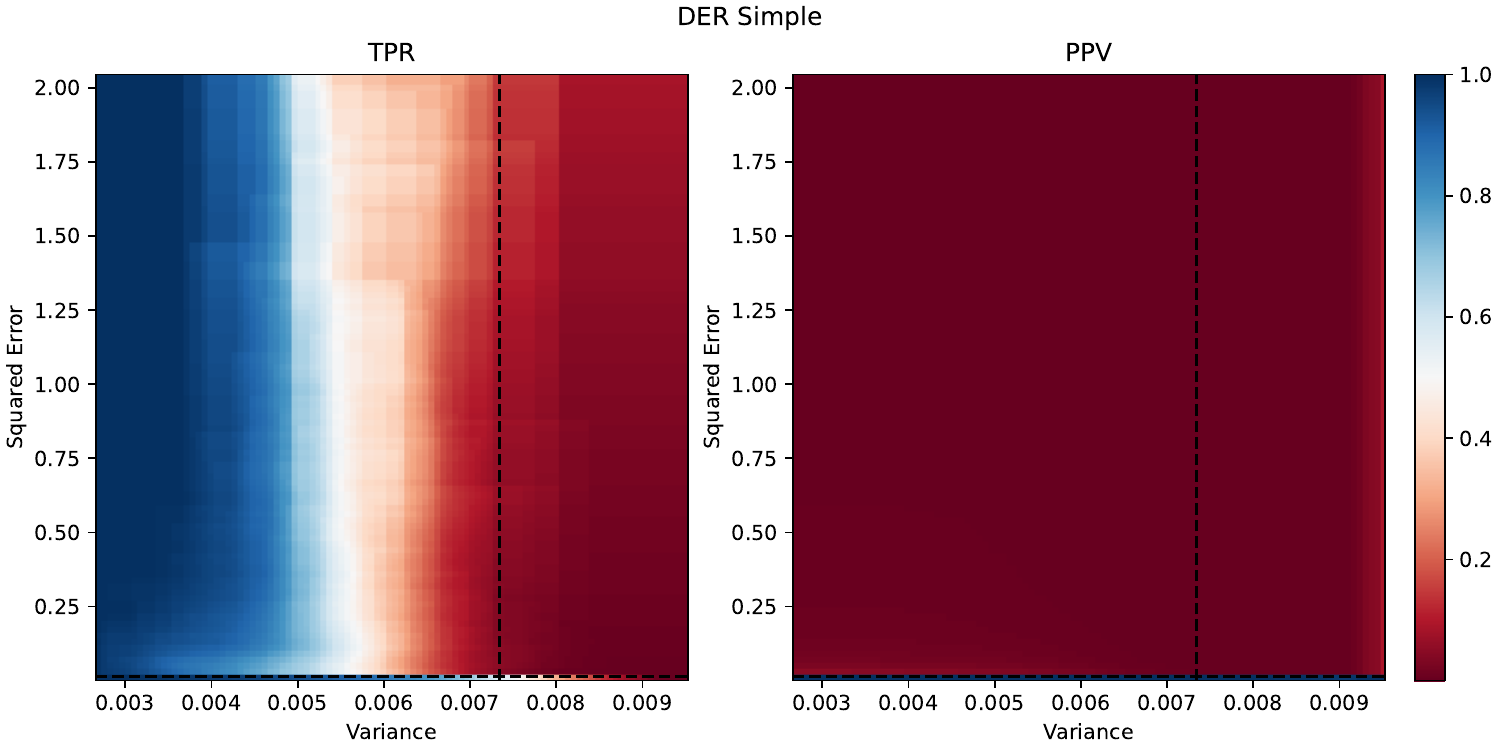}
    \caption{True Positive Rate (left) and Positive Predictive Value (right) for DER-S.}
    \label{sifig:tpr_der_simple}
\end{figure}

\begin{figure}
    \centering
    \includegraphics[width=0.9\textwidth]{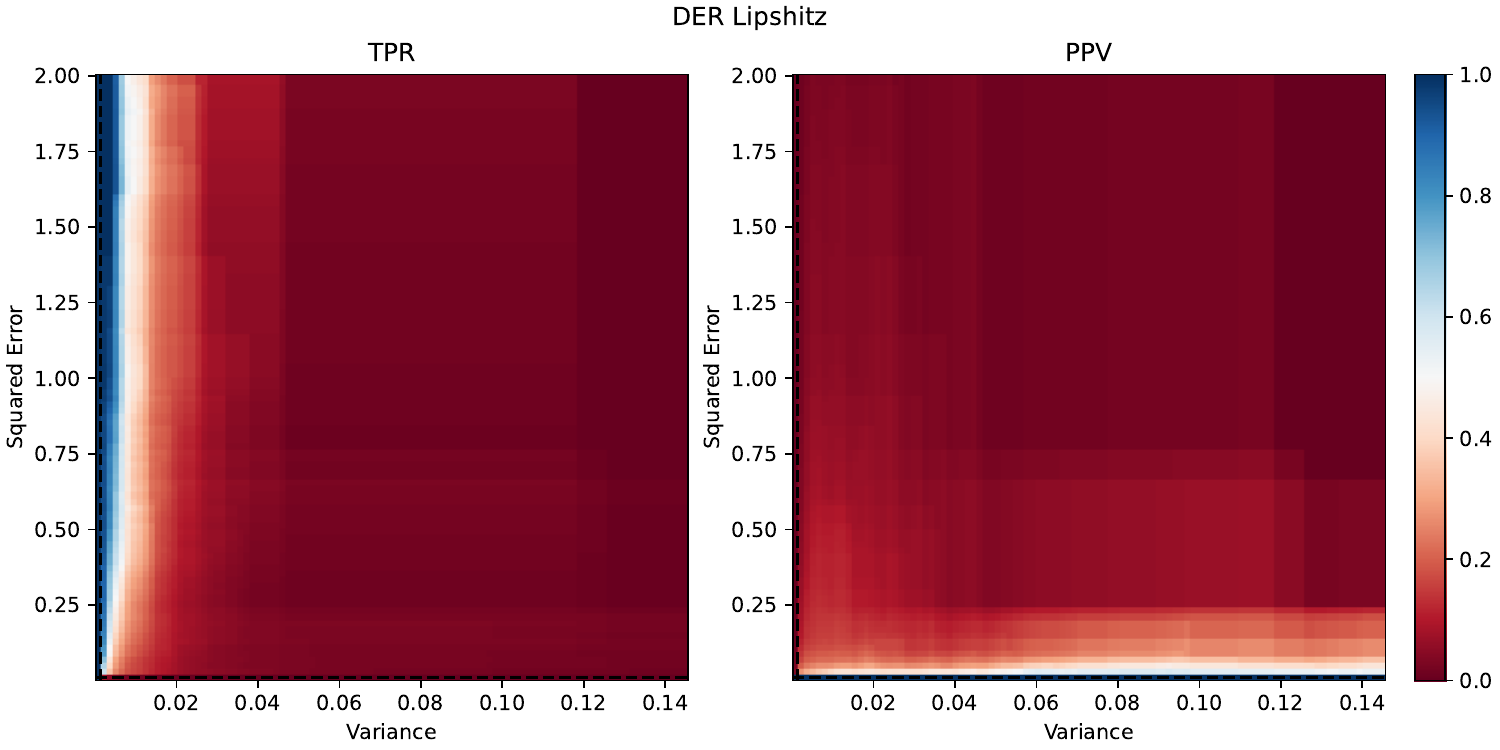}
    \caption{True Positive Rate (left) and Positive Predictive Value (right) for DER-L.}
    \label{sifig:tpr_lips}
\end{figure}

\begin{figure}
    \centering
    \includegraphics[width=0.9\textwidth]{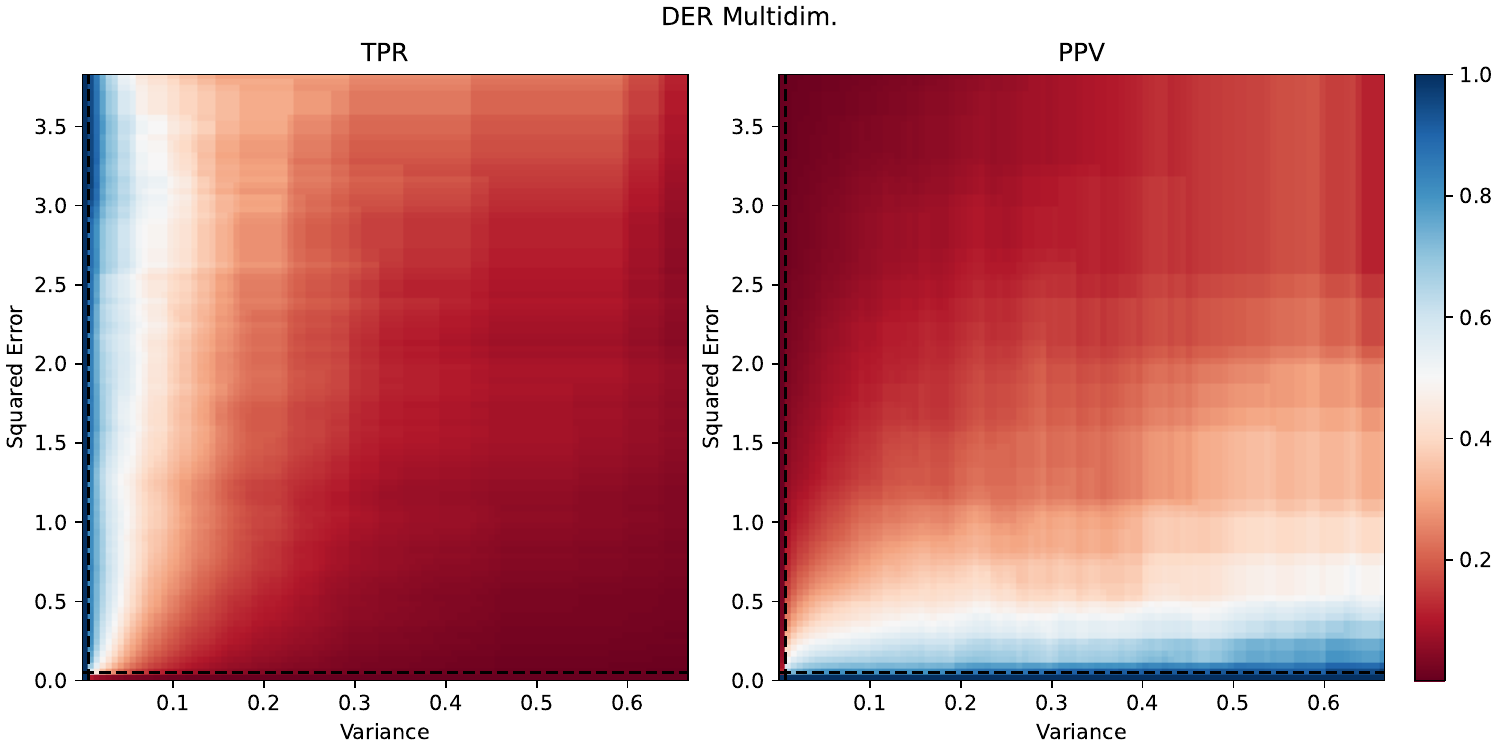}
    \caption{True Positive Rate (left) and Positive Predictive Value (right) for DER-M.}
    \label{sifig:tpr_MD}
\end{figure}

\begin{figure}
    \centering
    \includegraphics[width=0.9\textwidth]{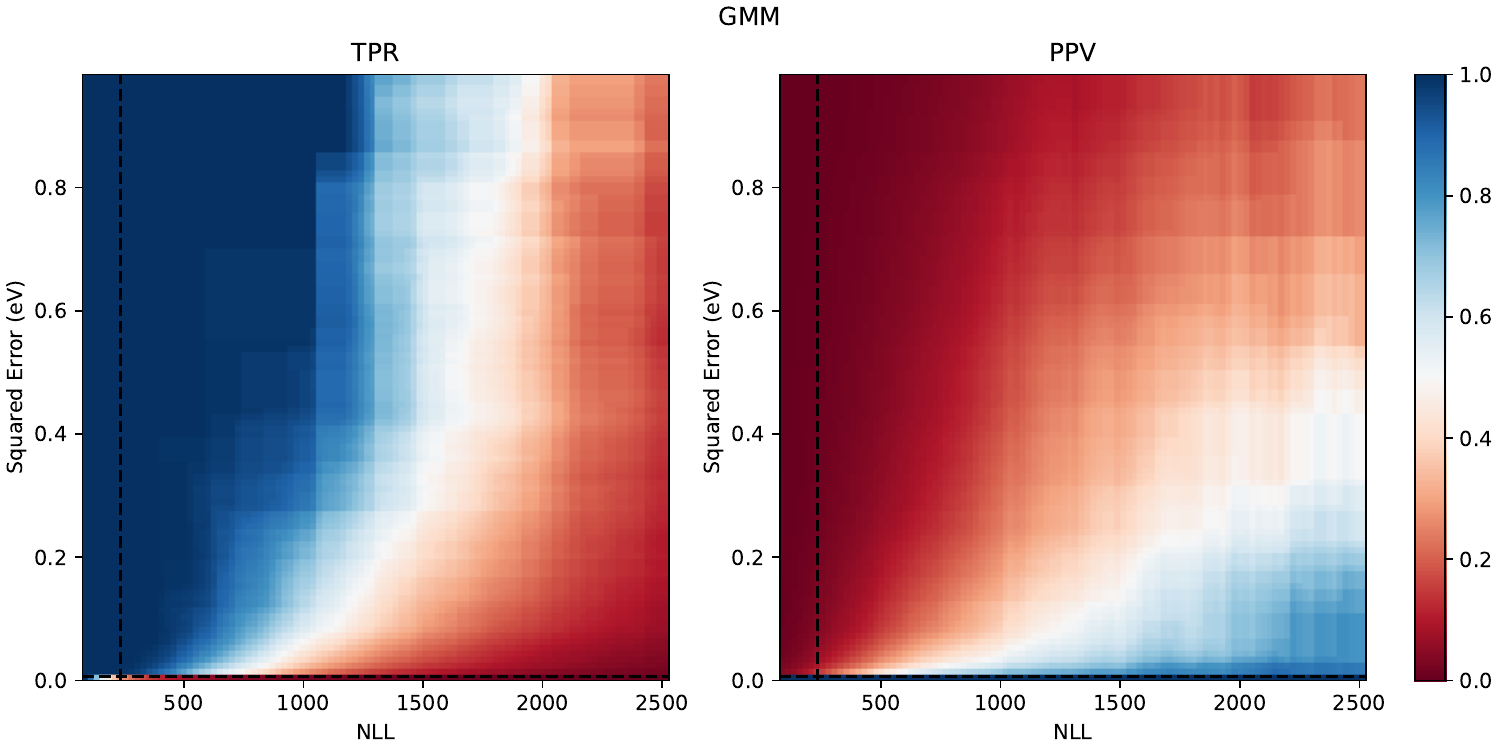}
    \caption{True Positive Rate (left) and Positive Predictive Value (right) for Gaussian Mixture Model.}
    \label{sifig:tpr_gmm}
\end{figure}

\begin{figure}
    \centering
    \includegraphics[width=0.9\textwidth]{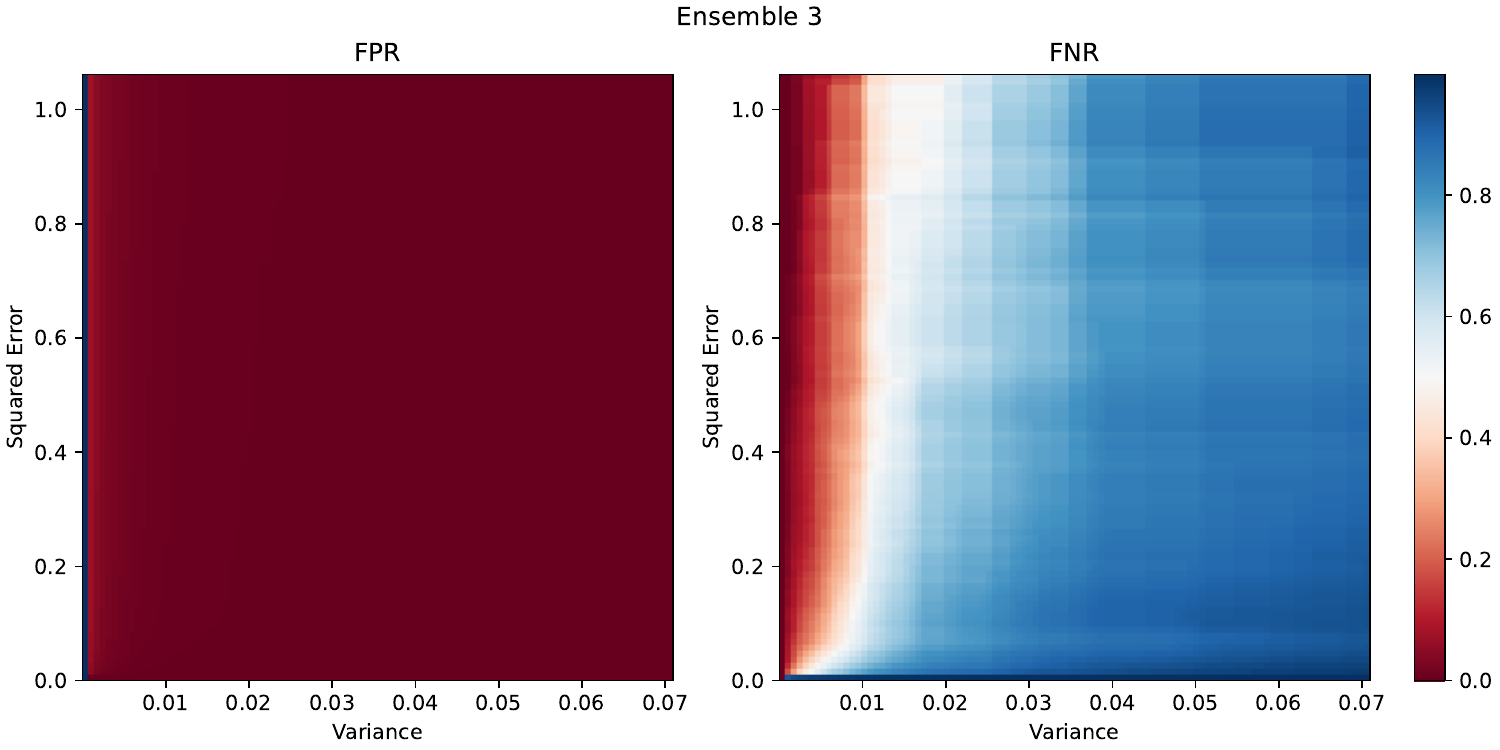}
    \caption{False Positive Rate (left) and False Negative Rate (right) for the Ens-3 model}
    \label{sifig:fpr_ens3}
\end{figure}

\begin{figure}
    \centering
    \includegraphics[width=0.9\textwidth]{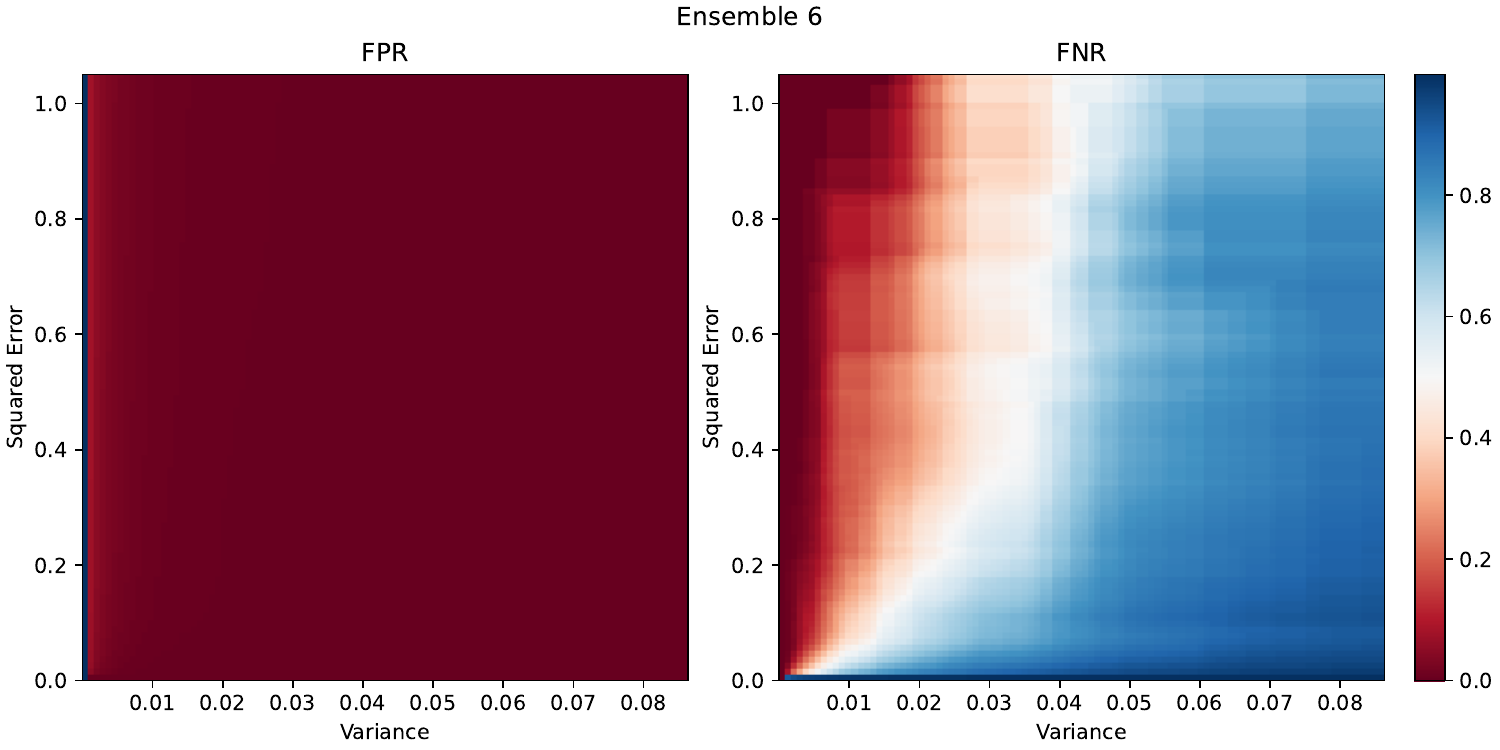}
    \caption{False Positive Rate (left) and False Negative Rate (right) for the Ens-6 model}
    \label{sifig:fpr_ens6}
\end{figure}

\begin{figure}
    \centering
    \includegraphics[width=0.9\textwidth]{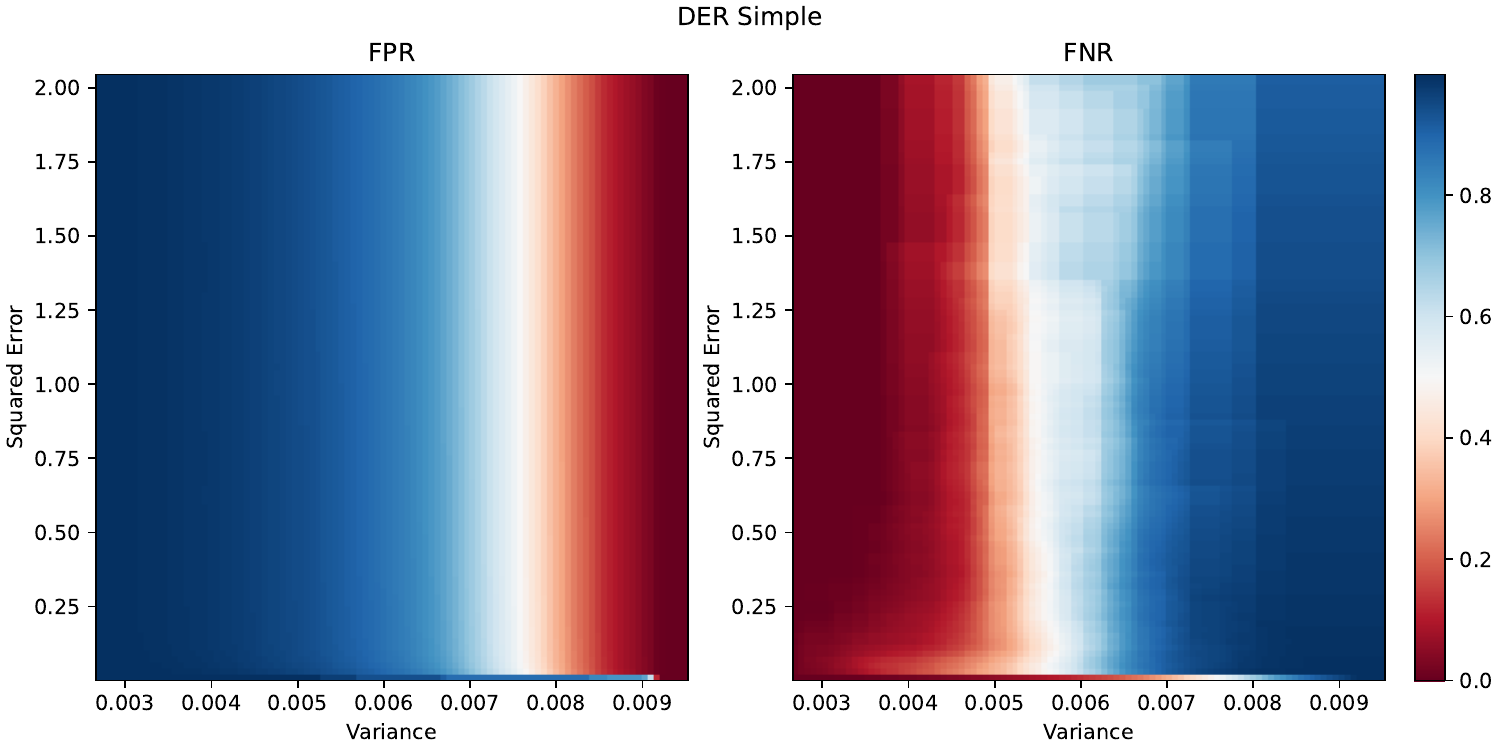}
    \caption{False Positive Rate (left) and False Negative Rate (right) for DER-S}
    \label{sifig:fpr_simple}
\end{figure}

\begin{figure}
    \centering
    \includegraphics[width=0.9\textwidth]{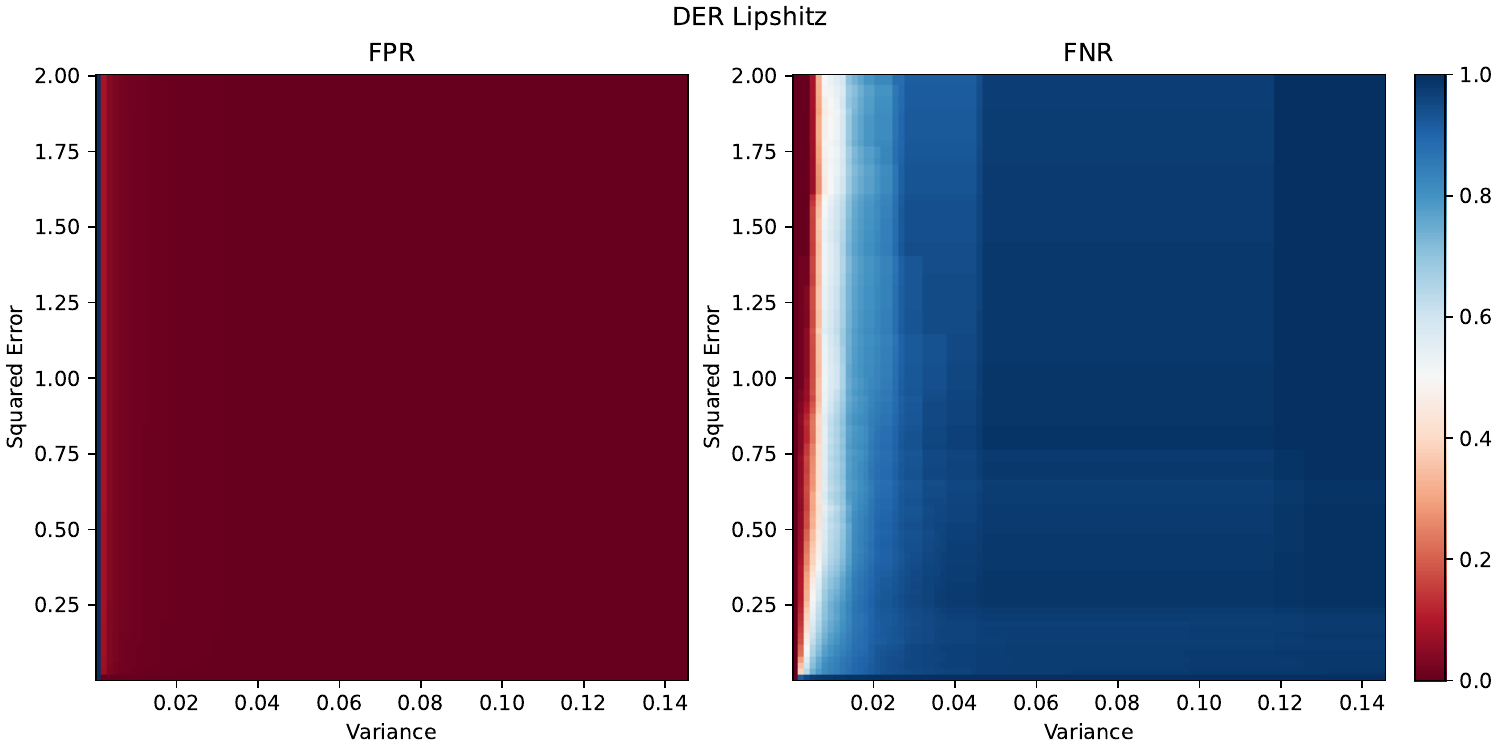}
    \caption{False Positive Rate (left) and False Negative Rate (right) for DER-L}
    \label{sifig:fpr_lips}
\end{figure}

\begin{figure}
    \centering
    \includegraphics[width=0.9\textwidth]{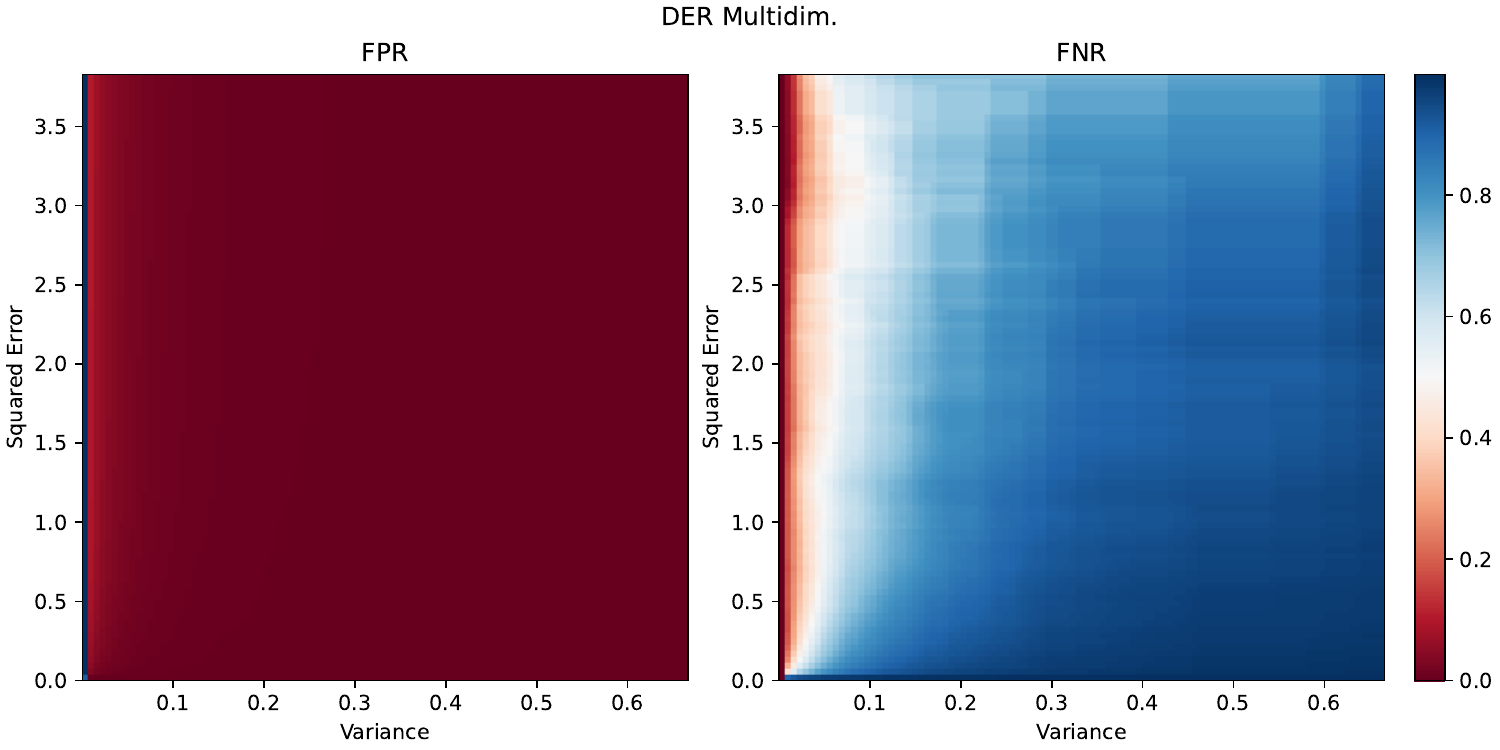}
    \caption{False Positive Rate (left) and False Negative Rate (right) for DER-M}
    \label{sifig:fpr_MD}
\end{figure}

\begin{figure}
    \centering
    \includegraphics[width=0.9\textwidth]{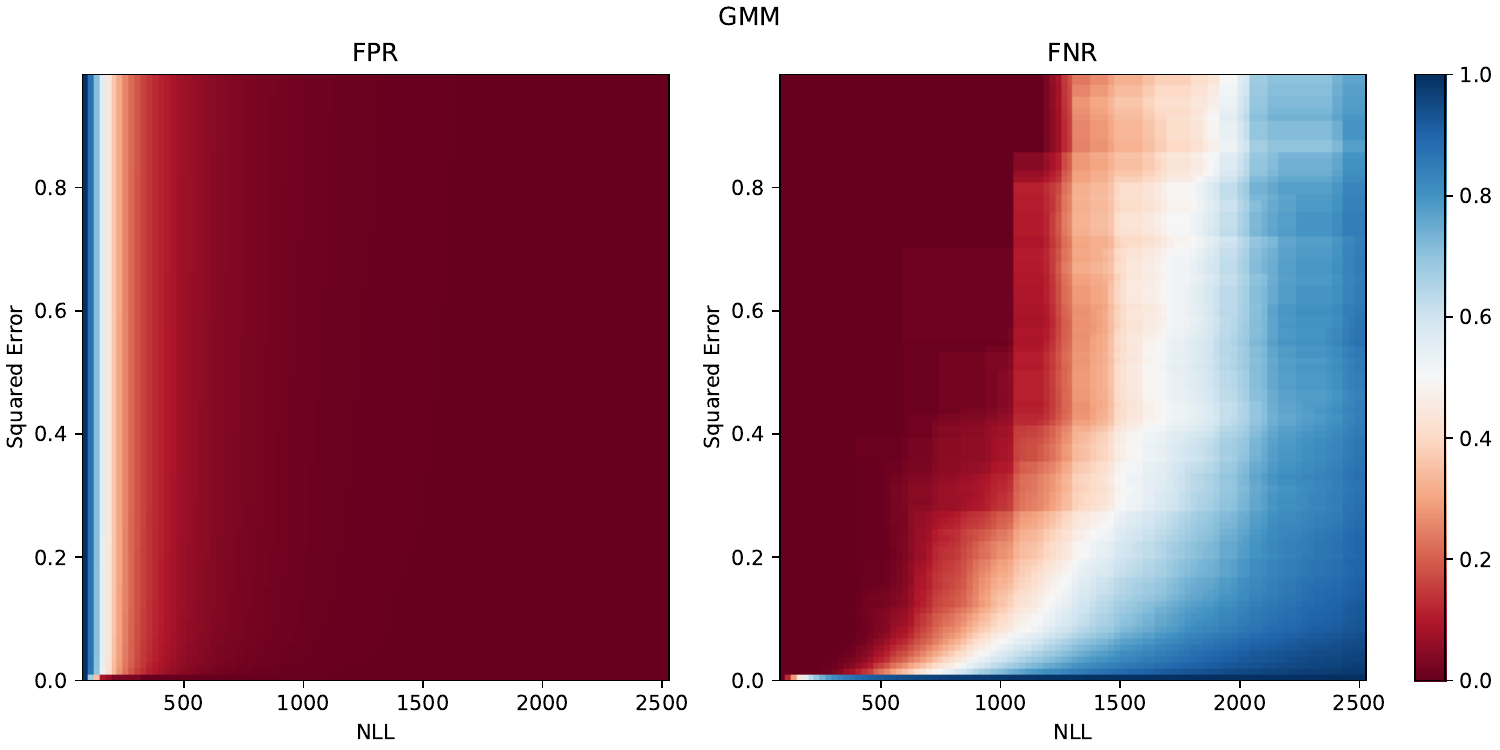}
    \caption{False Positive Rate (left) and False Negative Rate (right) for Gaussian Mixture Model}
    \label{sifig:fpr_gmm}
\end{figure}

\begin{figure}
    \centering
    \includegraphics[width=\textwidth]{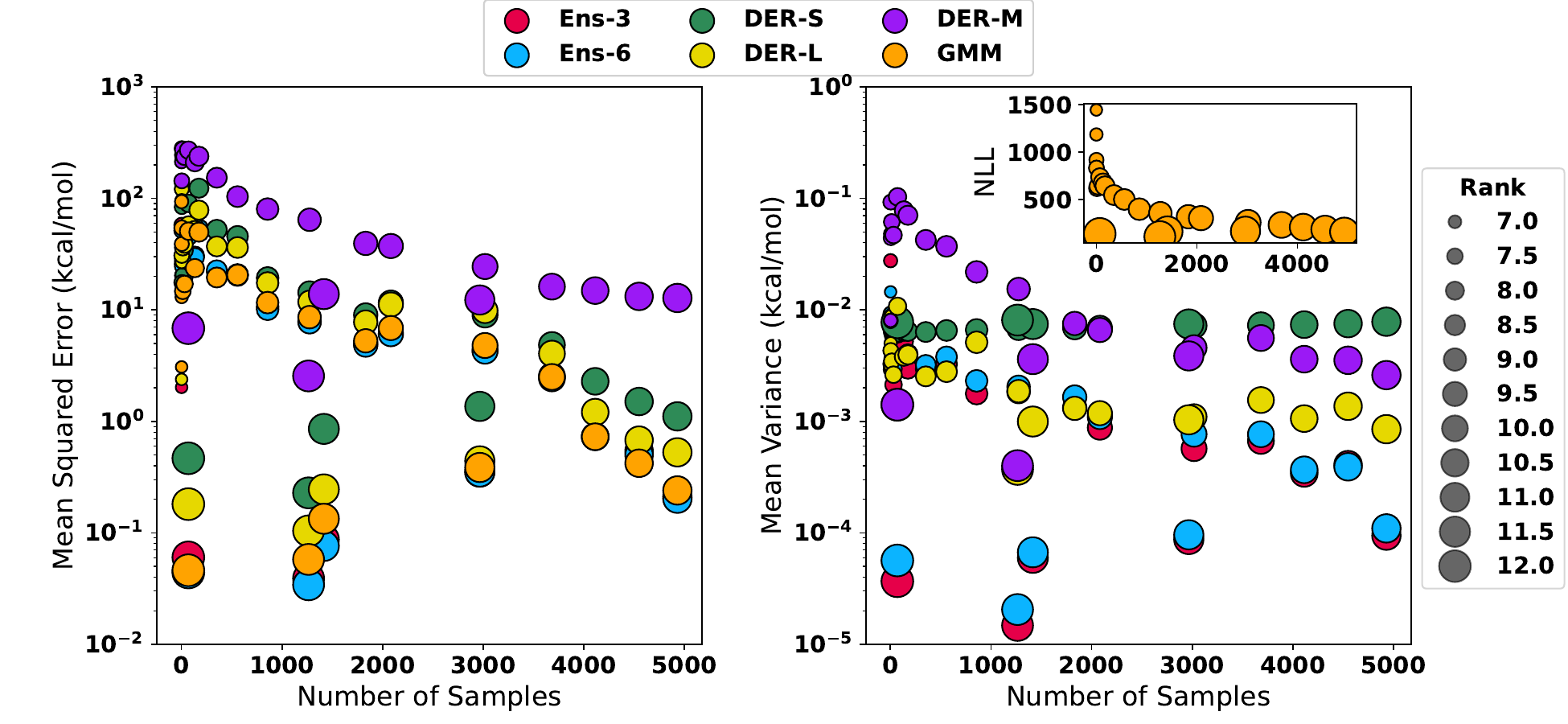}
    \caption{Changes in the mean square error (left) and mean variance (right) with respect to the number of samples in each class. The size of the scatter point is scaled with the ranking number. For the GMM model, the NLL is used to estimate the uncertainty. Notice that the $y$-axis scale is logarithmic.}
    \label{sifig:samples_v_mae}
\end{figure}

\begin{figure}
    \centering
    \includegraphics[scale=0.6]{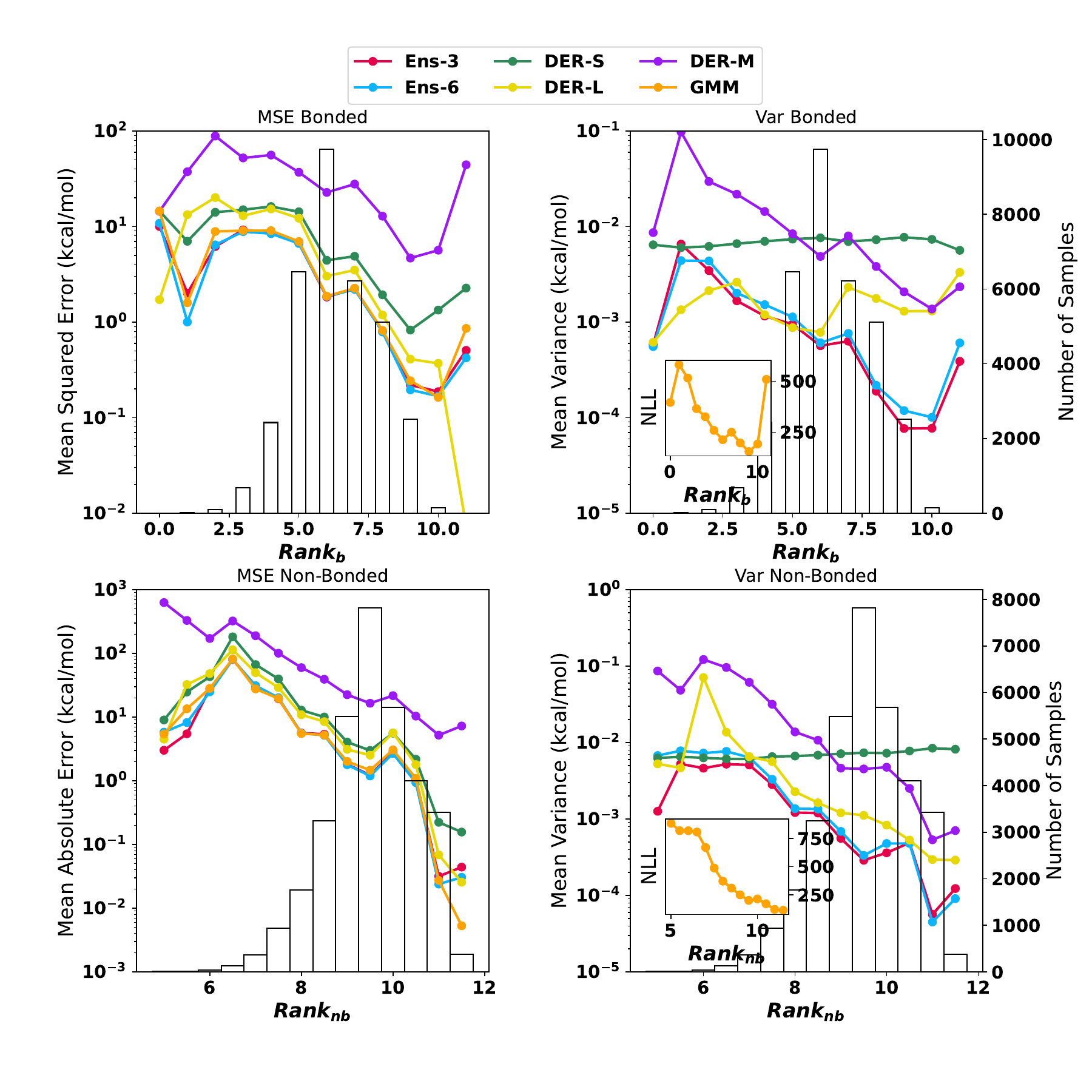}
    \caption{Changes in the mean square error (left) and mean variance (right) with respect to the rank of the molecules in the test set divided by contributions to bond (top) and non-bonded (bottom). In the background, a histogram of the number of samples with the same rank. For the GMM model, the NLL is used to estimate the uncertainty; therefore, the inset shows the changes in the NLL with respect to the rank. Notice that the $y$-axis scale is on logarithmic units.}
    \label{sifig:inout_bond_nobond}
\end{figure}

\begin{figure}
    \centering
    \includegraphics[scale=0.5]{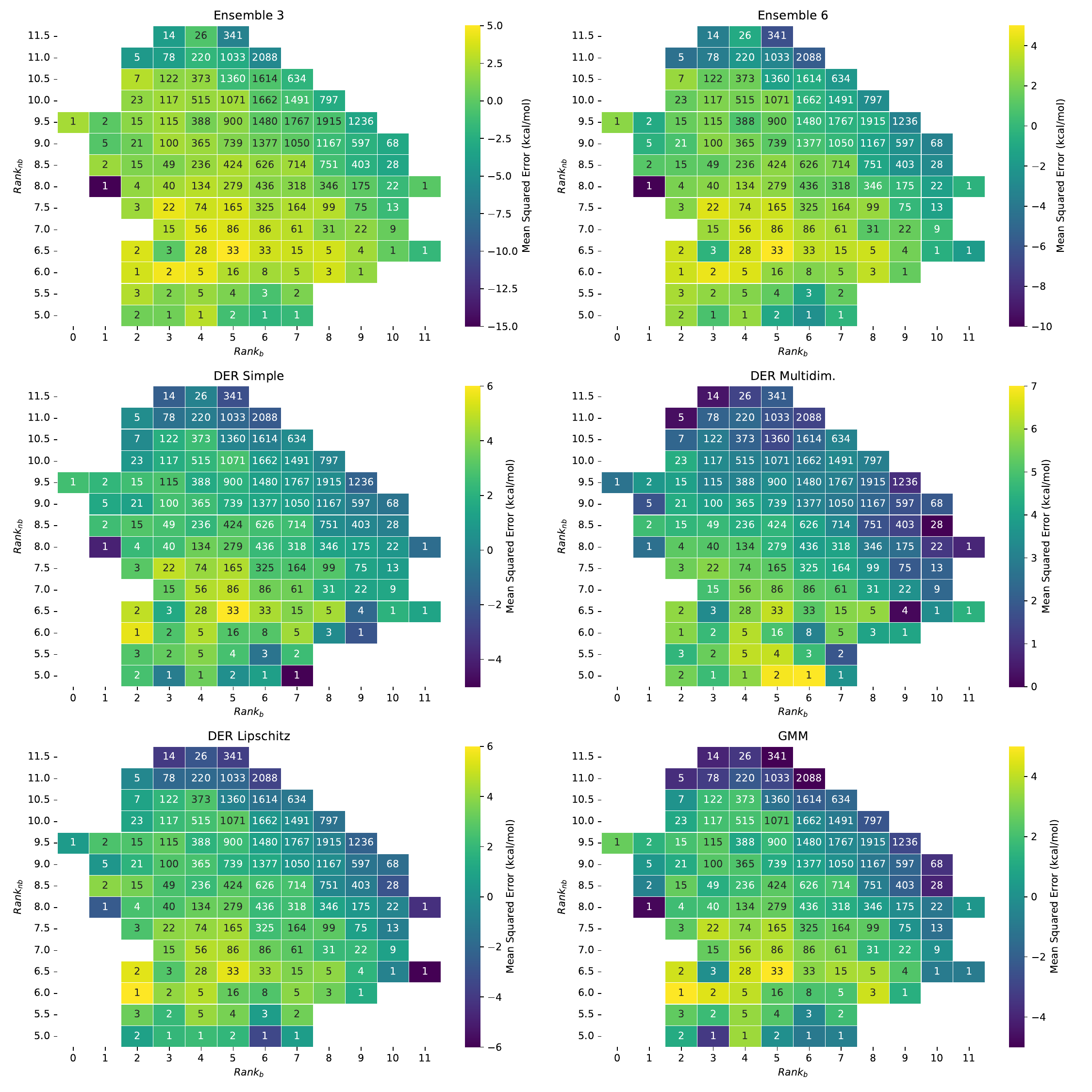}
    \caption{Map of influence of rank values for in and outside distribution of bond and non-bonded distances with respect to the error. The colour bar indicates the logarithm of the Mean Square Error and is normalised to its minimum and maximum values. The numbers inside each box are the number of samples for that score. The box is empty if no samples were found with that combination.}
    \label{sifig:mae_rank}
\end{figure}

\begin{figure}
    \centering
    \includegraphics[scale=0.475]{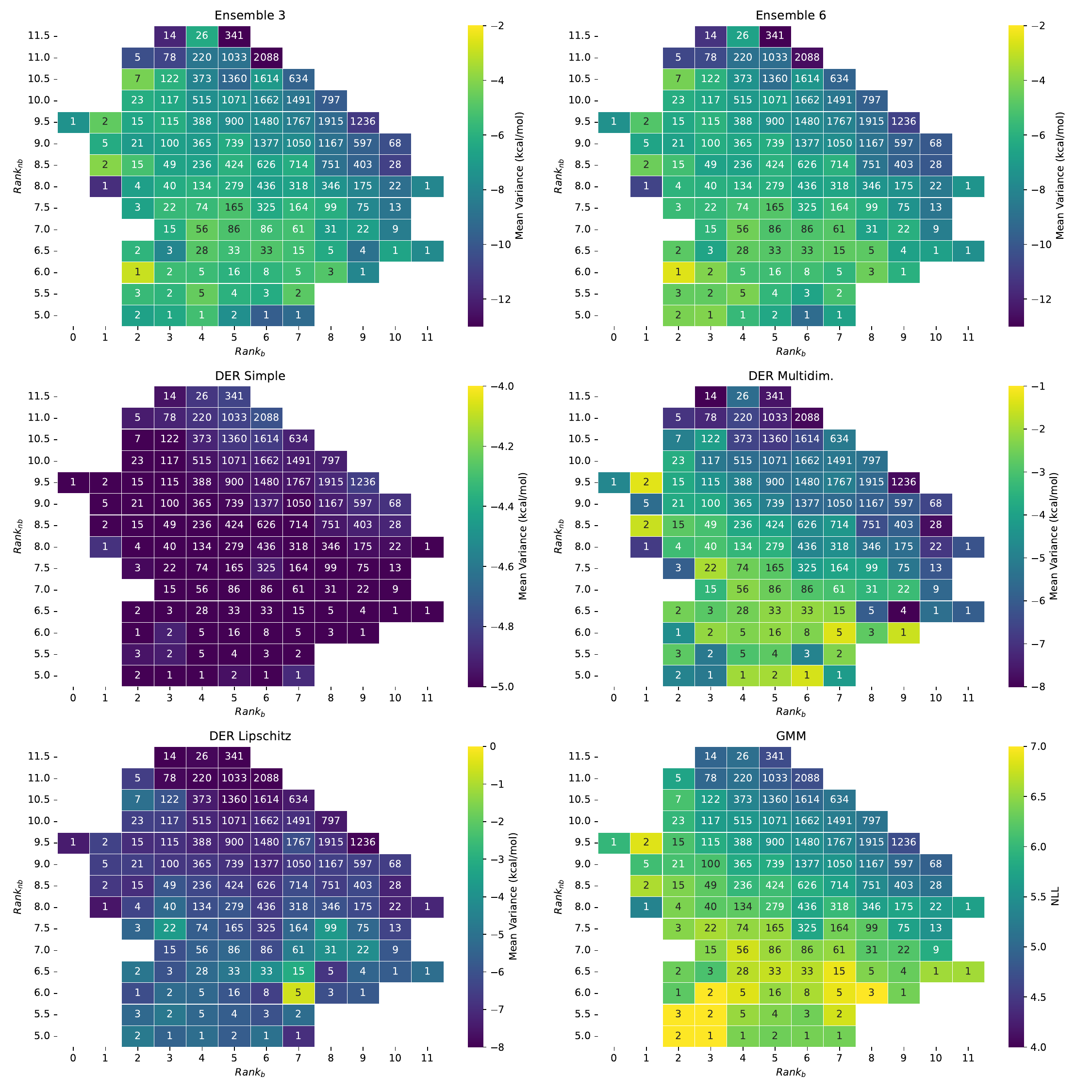}
    \caption{Map of influence of rank values for in and outside distribution of bond and non-bonded distances with respect to the variance. The colour bar indicates the logarithm of the mean variance except for GMM, which shows the NLL and is normalised to its minimum and maximum values. The numbers inside each box are the number of samples for that score. The box is empty if no samples were found with that combination.}
    \label{sifig:var_rank}
\end{figure}

\end{document}